\documentclass[12pt,a4paper]{article}

\usepackage[hyphens]{url}
\usepackage[top=2.5cm,bottom=3cm,left=2.5cm,right=2.5cm]{geometry}
\usepackage{mathtools,amssymb}
\usepackage[usenames,dvipsnames,table]{xcolor}
\usepackage[titletoc]{appendix}
\usepackage{mathtools,amssymb}
\usepackage[makeroom]{cancel}
\usepackage{graphicx}
\usepackage{epstopdf}
\usepackage{mathrsfs}
\usepackage{enumitem}
\usepackage{stackrel}
\usepackage{float}
\usepackage{booktabs}
\usepackage[normalem]{ulem}  
\usepackage{cancel} 
\usepackage{hyperref}
\usepackage[final,color]{showkeys} 
\usepackage{longtable}
\usepackage{physics}
\usepackage{empheq}
\usepackage{cite} 
\usepackage{breqn}

\usepackage{tikz}
\usetikzlibrary{intersections,shapes,arrows,positioning,automata,backgrounds,calc,er,patterns}
\tikzset{%
  link/.style    = { white, double = black, line width = 2.5pt,
                     double distance = 1.25pt },
  channel/.style = { white, double = black, line width = 5pt,
                     double distance = 1.25pt },
}
\usepackage[compat=1.1.0]{tikz-feynman}

\DeclareMathAlphabet{\mathpzc}{OT1}{pzc}{m}{it}

\allowdisplaybreaks
\allowbreak

\usepackage{etoolbox}
\makeatletter
\patchcmd{\hyper@makecurrent}{%
    \ifx\Hy@param\Hy@chapterstring
    \let\Hy@param\Hy@chapapp
    \fi
}{%
    \iftoggle{inappendix}{
	\@checkappendixparam{chapter}%
	\@checkappendixparam{section}%
	\@checkappendixparam{subsection}%
	\@checkappendixparam{subsubsection}%
	\@checkappendixparam{paragraph}%
	\@checkappendixparam{subparagraph}%
    }{}%
}{}{ \errmessage{failed to patch}}

\newcommand*{\@checkappendixparam}[1]{%
	\def\@checkappendixparamtmp{#1}%
	\ifx\Hy@param\@checkappendixparamtmp
	\let\Hy@param\Hy@appendixstring
	\fi
}
\makeatletter

\newtoggle{inappendix}
\togglefalse{inappendix}

\apptocmd{\appendix}{\toggletrue{inappendix}}{}{\errmessage{failed to patch}}
\apptocmd{\subappendices}{\toggletrue{inappendix}}{}{\errmessage{failed to patch}}

\begin{document}

\newcommand{\partiald}[2]{\dfrac{\partial #1}{\partial #2}}
\newcommand{\be}{\begin{equation}}
\newcommand{\ee}{\end{equation}}
\newcommand{\f}{\frac}
\newcommand{\s}{\sqrt}
\newcommand{\lm}{\mathcal{L}}
\newcommand{\wm}{\mathcal{W}}
\newcommand{\om}{\mathcal{O}_{n}}
\newcommand{\bea}{\begin{eqnarray}}
\newcommand{\eea}{\end{eqnarray}}
\newcommand{\ba}{\begin{align}}
\newcommand{\ea}{\end{align}}
\newcommand{\ep}{\epsilon}

\def\gap#1{\vspace{#1 ex}}
\def\be{\begin{equation}}
\def\ee{\end{equation}}
\def\bal{\begin{array}{l}}
\def\ba#1{\begin{array}{#1}}  
\def\ea{\end{array}}
\def\bea{\begin{eqnarray}}
\def\eea{\end{eqnarray}}
\def\beas{\begin{eqnarray*}}
\def\eeas{\end{eqnarray*}}
\def\del{\partial}
\def\eq#1{(\ref{#1})}
\def\fig#1{Fig \ref{#1}} 
\def\re#1{{\bf #1}}
\def\bull{$\bullet$}
\def\nn{\nonumber}
\def\ub{\underbar}
\def\nl{\hfill\break}
\def\ni{\noindent}
\def\bibi{\bibitem}
\def\vev#1{\langle #1 \rangle} 
\def\mattwo#1#2#3#4{\left(\begin{array}{cc}#1&#2\\#3&#4\end{array}\right)} 
\def\tgen#1{T^{#1}}
\def\half{\frac12}
\def\floor#1{{\lfloor #1 \rfloor}}
\def\ceil#1{{\lceil #1 \rceil}}

\def\Tr{{\rm Tr}}

\def\mysec#1{\gap1\ni{\bf #1}\gap1}
\def\mycap#1{\begin{quote}{\footnotesize #1}\end{quote}}

\def\Red#1{{\color{red}#1}}

\def\Om{\Omega}
\def\a{\alpha}
\def\b{\beta}
\def\l{\lambda}
\def\g{\gamma}
\def\e{\epsilon}
\def\Si{\Sigma}
\def\p{\phi}
\def\z{\zeta}

\def\lan{\langle}
\def\ran{\rangle}

\def\bit{\begin{item}}
\def\eit{\end{item}}
\def\benu{\begin{enumerate}}
\def\eenu{\end{enumerate}}

\def\tr{{\rm tr}}
\def\intk#1{{\int\kern-#1pt}}
\def\tsum{\textstyle \sum}

\def\al{\alpha}
\def\ga{\gamma}
\def\Ga{\Gamma}
\def\G{\Gamma}
\def\be{\beta}
\def\de{\delta}
\def\De{\Delta}
\def\ep{\epsilon}
\def\ro{\rho}
\def\la{\lambda}
\def\La{\Lambda}
\def\ka{\kappa}
\def\om{\omega}
\def\si{\sigma}
\def\th{\theta}
\def\ze{\zeta}
\def\ne{\eta}
\def\del{\partial}
\def\cdev{\nabla}

\def\gh{\hat{g}}
\def\Rh{\hat{R}}
\def\Boxh{\hat{\Box}}
\def\Kb{\mathcal{K}}
\def\phit{\tilde{\phi}}
\def\gt{\tilde{g}}
\newcommand{\h}{\hat}
\newcommand{\ti}{\tilde}
\newcommand{\sD}{{\mathcal{D}}}
\newcommand{\colored}[1]{ {\color{turquoise} #1 } }
\newcommand{\propBbd}{\mathcal{G}}
\newcommand{\propBB}{\mathbb{G}}
\newcommand{\christof}[3]{ {\Ga^{#1}}_{#2 #3}}
\def\ads{AdS$_{\text{2}}$~}
\def\GN{G$_{\text{N}}$~}
\def\zb{{\bar{z}}}
\def\fb{\bar{f}}
\def\delb{\bar{\del}}
\def\wb{\bar{w}}
\def\gb{\bar{g}}
\def\gp{g_+}
\def\gm{g_-}
\def\phit{\tilde{\phi}}
\def\mut{\tilde{\mu}}
\def\xb{\bar{x}}
\def\yb{\bar{y}}
\def\xp{x_+}
\def\xm{x_-}
\def\finv{\mathfrak{f}_i}
\def\fbinv{\bar{\mathfrak{f}}_i}
\def\gc{\mathfrak{g}}
\def\gcb{\bar{ \mathfrak{g}}}
\def\disc{\mathcal{D}}
\def\rhp{\mathbb{H}}
\def\picklemma{Schwarz-Pick lemma}
\def\mobius{M\"{o}bius~}
\def\ft{\tilde{f}}
\def\zet{\tilde{\ze}}
\def\taut{\tilde{\tau}}
\def\thet{\tilde{\theta}}
\def\slr{\ensuremath{\mathbb{SL}(2,\mathbb{R})}}
\def\slc{\ensuremath{\mathbb{SL}(2,\mathbb{C})}}
\def\nh{\hat{n}}
\def\cD{\mathcal{D}}
\def\Lfg{\mathfrak{f}}

\def\va{{\bf a}}
\def\vb{{\bf b}}
\def\vc{{\bf c}}
\def\vd{{\bf d}}
\def\ve{{\bf e}}
\def\vf{{\bf f}}
\def\vg{{\bf g}}
\def\vh{{\bf h}}
\def\vr{{\mathbf{r}}}
\def\vx{{\mathbf{x}}}
\def\vF{{\mathbf{F}}}
\def\vB{{\mathbf{B}}}
\def\vp{{\mathbf{p}}}
\def\vq{{\mathbf{q}}}
\def\vk{{\mathbf{k}}}
\def\vs{{\mathbf{s}}}

\def\vtp{{\bf \tilde{p}}}
\def\vtq{{\bf \tilde{q}}}
\def\tp{{ \tilde{p}}}
\def\tq{{ \tilde{q}}} 
\def\vpp{{\mathbf{ P}}}
\def\vtpp{{\bf \tilde{ P}}}
\def\dt{\widetilde{dt}}
\def\id{\mathds{1}}
\def\cB{\mathcal{B}}
\def\cU{\mathcal{U}}
\def\cN{\mathcal{N}}
\def\cK{\mathcal{K}}
\def\cR{\mathcal{R}}
\def\cO{\mathcal{O}}
\def\cH{\mathcal{H}}
\def\cF{\mathcal{F}}

\renewcommand{\real}{\mathbb{R}}

\newcommand*{\Cdot}[1][1.25]{%
  \mathpalette{\CdotAux{#1}}\cdot%
}
\newdimen\CdotAxis
\newcommand*{\CdotAux}[3]{%
    {%
	\settoheight\CdotAxis{$#2\vcenter{}$}%
	\sbox0{%
	    \raisebox\CdotAxis{%
		\scalebox{#1}{%
		    \raisebox{-\CdotAxis}{%
			$\mathsurround=0pt #2#3$%
		    }%
		}%
	    }%
	}%
	\dp0=0pt %
	\sbox2{$#2\bullet$}%
	\ifdim\ht2<\ht0 %
	\ht0=\ht2 %
	\fi
	\sbox2{$\mathsurround=0pt #2#3$}%
	\hbox to \wd2{\hss\usebox{0}\hss}%
    }%
}

%

\def\newthing{\marginpar{{\color{red}****}}}
\reversemarginpar
\def\bz{{\bar z}}
\def\by{{\bar y}}
\def\bw{{\bar w}}
\def\delb{{\bar \del}}
\def\bep{{\bar \epsilon}}

\hypersetup{pageanchor=false}

\begin{titlepage}
\begin{flushright}
\end{flushright}

\bigskip

    \begin{center}
	{\Large \bf{Non-Gaussian Entanglement Renormalization 
	\\[15pt]	
	for Quantum Fields 	
	}}
	
	\vskip 2cm

	{\bf 	J.J. Fern\'andez-Melgarejo$^{1}$ \  and \  J. Molina-Vilaplana$^{2}$}

	\begin{center}
	    {\it $^{1}$\, Departamento de F\'isica, Universidad de Murcia, Spain}\\
	     {\it  $^{2}$\, Universidad Polit\'ecnica de Cartagena, Spain}\\	    	    
	\end{center}

	\gap2


    \end{center}
    
\vskip 1.5cm

  \begin{abstract}
In this work, a non-Gaussian cMERA tensor network for interacting quantum field theories (icMERA) is presented. This consists of a continuous tensor network circuit in which the generator of the entanglement renormalization of the wavefunction is nonperturbatively extended with nonquadratic variational terms. The icMERA circuit nonperturbatively implements a set of scale dependent nonlinear transformations on the fields of the theory, which suppose a generalization of the scale dependent linear transformations induced by the Gaussian cMERA circuit. Here we present these transformations for the case of self-interacting scalar and fermionic field theories. Finally, the icMERA tensor network is fully optimized for the $\lambda \phi^4$ theory in $(1+1)$ dimensions. This allows us to evaluate, nonperturbatively, the connected parts of the two- and four-point correlation functions. Our results show that icMERA wavefunctionals encode proper non-Gaussian correlations of the theory, thus providing a new variational tool to study phenomena related with strongly interacting field theories.  
\end{abstract}

\end{titlepage}

\pagenumbering{roman}
\tableofcontents
\enlargethispage{1000pt}
\pagebreak
\pagenumbering{arabic}
\setcounter{page}{1}

\section{Introduction and Summary}\label{sec:Intro}

The multiscale entanglement renormalization ansatz (MERA) \cite{MERA1, MERA2}, which was originally proposed as a variational method to obtain the ground state of spin chains systems, consists of a real space renormalization group technique that, iteratively,  removes the quantum correlations between small adjacent regions of space at each length scale. A continuous version of MERA (cMERA) was proposed for free field theories \cite{cMERA1, cMERA2}. Motivated, among others, by the conjecture that cMERA is a realization of the AdS/CFT correspondence \cite{HolCMERA0, HolCMERA1,HolCMERA2,HolCMERA3,HolCMERA4, JMV1, JMV2}, a rigorous and (non)perturbative formalism for interacting theories turns out to be essential to advance in this program.

\medskip

Precisely, one of the major problems in quantum field theory (QFT) is the understanding of the phenomena associated to strongly coupled systems. To do so, nonperturbative methods are required. While these are difficult problems to solve exactly, it is acknowledged that the use of nonperturbative variational methods allow to tackle these problems to some extent. Despite field theory was initially formulated within the Hamiltonian framework, these methods lost relevance against path integral techniques for many years. Nevertheless, when it comes to nonperturbative aspects, there are situations in which the use of wavefunctionals on configuration space exhibits clear advantages. Namely, nonperturbative path integral methods are especially suited to compute quantities that have no perturbative contributions and can be addressed through a saddle point approximation. However, when the observables of interest receive both perturbative and nonperturbative contributions, dealing with the path integral becomes more difficult. In addition, with the raising appeal in understanding QFT from a quantum information point of view, the Hamiltonian framework involving wavefunctionals seems more suitable than other approaches.

\medskip

In recent years, tensor networks, a new class of variational states, have proven to be very useful in the study of a huge variety of interacting many body systems. Initially devised for lattice systems, through the Rayleigh-Ritz variational principle, a tensor network representation of the wavefunction provides an efficient approximation to the ground state of an interacting many body system by systematically identifying the relevant degrees of freedom for the physics at low energies. As an instance, the MERA tensor network \cite{MERA1, MERA2}, implements a variational real space renormalization group on the wavefunction that represents the ground state of the system at different length scales. In spite of their success in analyzing 1D systems on the lattice, several difficulties arise when trying to generalize tensor networks to higher dimensions and/or to interacting field theories. In this sense, the continuous generalization of the the matrix product state (cMPS), while proving efficient to describe the low energy physics of nonrelativistic systems in 1+1 dimensions, suffers from regularization ambiguities when dealing with relativistic systems in 1+1 dimensions \cite{stojevic15}.

\medskip

A continuous tensor network circuit designed to work in arbitrary spatial dimensions for (non)relativistic field theories is the continuous version of MERA, cMERA \cite{cMERA1, cMERA2}.  This tensor network builds a multi-layered representation of the ground state wavefunctional through a variationally optimized pattern of entanglement between the relevant degrees of freedom at any length scale. cMERA amounts to a real space renormalization group (RG)  of the wavefunctionals in a Hamiltonian framework  such that, each layer of the network corresponds to a step of the RG flow. Regrettably, cMERA has only been explicitly formulated for free theories of bosonic, fermionic and gauge fields \cite{cMERA1,cMERA2,cMERA3}. In these formulations, the cMERA renormalization in scale is generated by a quadratic operator, and thus, the resulting state is given by a Gaussian wavefunctional. Obviously, while this fact dramatically limits the interest of the Gaussian cMERA \emph{ansatz} for interacting QFTs, such trial states result useful to reproduce correlation functions and entanglement entropy in free field theories \cite{cMERA4}. However, to clarify whether cMERA is a possible realization of the holographic duality \cite{HolCMERA0, HolCMERA1,HolCMERA2,HolCMERA3,HolCMERA4, JMV1, JMV2}, a more general formulation to study nonperturbative interacting field theories becomes crucial. 

\medskip

Our aim in this work is to develop a truly non-Gaussian cMERA tensor network circuit able to nonperturbatively capture relevant phenomena of interacting QFTs. In this respect, let us comment on some aspects that stem from applying a variational method (\textit{e.g.}, tensor networks) to QFT: generality, calculability and ultraviolet modes.

\medskip

Firstly, one has the problem of the \emph{generality} of the trial state. Namely, the trial state should be general enough to capture the most salient physical features of the phenomena under study through the variation of its parameters. Due to the enormous size of the Hilbert space in a QFT, it is very difficult to identify by mere intuition the relevant parameters that have to be probed. Thus, a systematic method to build \emph{ansatze} is desirable. 

\medskip

Secondly, there is the problem of \emph{calculability} \cite{kovner}. That is to say, even when possessing a reasonable and flexible \emph{ansatz} for the vacuum wavefunctional, one needs to evaluate expectation values of operators/observables of interest in this state, 
\begin{equation}
\langle \cO \rangle=\int {\rm D}\phi \Psi^*[\phi]\, \cO\, \Psi[\phi]\, .
\end{equation}
This amounts to the evaluation of an Euclidean functional integral in which the square of the wavefunctional acts as the partition function.  Thus, in QFT, given the very limited ability to evaluate non-Gaussian path integrals, the calculability requirement on the trial wave functional is certainly severe. Indeed, it is so severe that it has constrained the form of the trial wavefunctionals to Gaussian states. Despite this, Gaussian trial states have been successfully applied to self-interacting relativistic scalar and spinor field theories \cite{moshe}, where a great amount of nontrivial exact results in the large $N$ limit has been obtained (among others, a proposal to build a Gaussian approach to cMERA for interacting field theories \cite{Cotler1}). Namely, the Gaussian \emph{ansatz} (that is the exact ground state in noninteracting QFTs) works very well for settings in which the relevant  nonperturbative physics  of the system is dominated by a single condensate. However, it is well known that the connected part of $N$-point correlation functions distinguishes the ground states of interacting theories from those of noninteracting ones. While for Gaussian states, the connected correlation functions of order higher than two vanish, those of interacting systems are generally nonzero. For this reason, despite being successful in capturing some nonperturbative effects, with the aim of going beyond the Gaussian approach, it would be desirable to develop in a systematic way, \emph{ansatze} which variationally borne in, some kind of ``generalized'' condensates while keeping the calculability of the Gaussian \emph{ansatz} intact.

\medskip

Finally, but not least important, one is faced to the problem of the  ultraviolet modes. The main objective of a variational calculation in a strongly interacting field theory is to obtain the correct configuration for the low momentum modes of the field in the vacuum wavefunctional. Due to the interaction between the high and low momentum modes in an interacting QFT, it is thus desirable to have a method that yields  variational parameters that optimally integrate out the effects of high energy modes into the low energy physics.

\medskip

Bearing in mind these features, various proposals have been recently raised to go beyond a purely Gaussian wavefunctional. In \cite{Cotler2, Cotler3}, authors have developed techniques to carry out systematic perturbative calculations of cMERA circuits restricted to a weakly interacting regime.  Another recent approach, \cite{magicmera}, proposes a particular realization of a Gaussian cMERA with an UV structure analogous to that of the cMPS. It is expected that this connection could yield cMERA wavefunctionals that are able to capture nonperturbative physics beyond the Gaussian approximation. In \cite{JJ1}, authors presented a truly nonperturbative method to build non-Gaussian cMERA wavefunctionals for interacting QFTs. This approach relies on nonlinear canonical transformations (NLCT) \cite{polley89, ritschel90, ritschel94,ibanez} to build a set of scale dependent extensive wavefunctionals which are certainly non-Gaussian. With this prescription, we showed that observables such as disconnected correlation functions can be analytically calculated in a closed form. 

\medskip

Let us remark the last proposal. The scale dependent NLCT in \cite{JJ1}, which were constructed from the product of two unitary operators, obviously admit a realization through a unitary operator. This is what could constitute a more satisfactory and natural continuum generalization of the lattice MERA algorithms. In the present work we refine our method and solve this issue using the firm ground provided by the conceptual framework presented in \cite{JJ1}. In addition, we carry out an exhaustive optimization process and show the efficiency and predictability of our method.

\medskip

Let us comment on the structure of the present work and summarize the main results of each section. In Section \ref{sec:formalism} we have reviewed and generalized the cMERA formalism for any entangler beyond the Gaussian one, which is a quadratic operator. Using a quantum mechanical toy model, 
one can prove that only when commuting with order-2 operators, an order-$k$ operator does not increase the order (see \eqref{eq:order-commutator}, \cite{Cotler2, Cotler3}):
\begin{align}
[\text{order }2,\ \text{order }k]\le \text{order } k
\ .
\label{eq:order-commutator-intro}
\end{align}
On the other hand, entanglers containing order-$k$ operators, $k\ge3$, are necessary to generate non-Gaussian states. However, the Hadamard's lemma and the above result show that they are very problematic when acting on the field operators of the theory, which are typically of the form $\phi^2$, $\pi^2$ or $\phi^n$.  
Thus, if the approach must be nonperturbative, an alternative solution appears to be necessary to solve this puzzle. This is what we will propose in Sections \ref{sec:ng-cMERA} and \ref{sec:i_cMERA_circuit}.

\medskip

In Section \ref{sec:gaussian} we have reviewed the quadratic Gaussian entangler introduced in \cite{cMERA1}, which performs scale dependent Bogoliubov transformations on the fields. Based on \eqref{eq:order-commutator-intro}, we can grasp why both free and interacting theories can be treated with Gaussian trial wavefunctions: when applying the Gaussian entangler, the order of the renormalized operators does not grow. Afterward, upon studying the free and the $\lambda\phi^4$ scalar theory, we give evidence of the limitations of the Gaussian entangler to study interacting theories. For example, we introduce the connected $N$-point correlators and justify why the 4-point correlators or the kurtosis automatically vanish for Gaussian trial wavefunctions.

\medskip

In Section \ref{sec:ng-cMERA} we have firstly studied the nonquadratic operator \eqref{eq:BBos}
\begin{align}\label{eq:BBos-intro}
\mathcal{B}
= 
-s\int_{\vp \vq_1\cdots \vq_m} 
h(p,q_1,\ldots,q_m)\, \pi(\vp)\, \phi(\vq_1)\ldots\phi(\vq_m) \delta(\vp+\vq_1+\cdots\vq_m) 
\ ,
\end{align}
which generates NLCTs on the field operators. Being a nonperturbative operator, we have studied the non-Gaussian variational parameters and given the prescription for which the series of nested commutators truncates. Secondly, we have reviewed \cite{JJ1} as a primary method to introduce the scale dependence on the NLCT. This is a key construction for the final formulation of the circuit. Finally, we have proposed a unitary operator that performs a set of NLCTs on chiral Dirac fields in 1+1 dimensions. Such unitary is written in terms of the Hermitian operator \eqref{eq:fermionic_transform}
\begin{eqnarray}\label{eq:fermionic_transform-intro}
	\cF
	&=&s\, \int_{\vq_1\cdots\vq_4}g(\vq_1,\cdots,\vq_4)\bar\psi_R(\vq_1)\gamma^0\psi_L(\vq_2)\bar\psi_R(\vq_3)\gamma^0\psi_L(\vq_4)\delta(\tsum_i\vq_i)
	\ .
\end{eqnarray}
Naturally, $\cF$ will play the role of a fermionic entangler when scale dependence is included.

\medskip

In Section \ref{sec:i_cMERA_circuit}, we have considered a circuit formed by the exponential of the entangler $K(u)=K_0(u)+\cB(u)$, where $K_0(u)$ is the Gaussian entangler and $\cB(u)$ is
\begin{align}\label{eq:icMERA_dis-intro}
\cB(u)
=
s \int_{\vq_1\vq_2\vq_3} g(q_1,q_2,q_3;u) \pi(\vq_1)\phi(\vq_2)\phi(\vq_3)\delta(\vq_1+\vq_2+\vq_3)
\ .
\end{align}
Afterward, to evaluate the quality of our scale dependent non-Gaussian ansatz, various observables have been studied. Due to their versatility and usefulness, we have obtained in a closed and analytical form the connected $N$-point correlation functions, $N\le4$. Then, upon considering the $\lambda \phi^4$ theory, we have carried out the full optimization of the icMERA circuit corresponding to the ground state of the theory. The optimization yields closed analytical expressions for the variational scale dependent parameters of the system. Subsequently, we have evaluated the 2-point and 4-point connected correlators with the optimized variational parameters at various scales. Interestingly, the connected 4-point correlators exhibit an unambiguous non-Gaussian behavior at various scales.

\medskip

Finally, Section \ref{sec:discussion} contains the conclusions extracted from the results, together with some open questions that are expected to be relevant for this research program.

\medskip

Appendix \ref{app:diagrams} provides a perturbative analysis of icMERA for a particular model, whereas Appendix \ref{app:integrals} contains a set of scale dependent loop integrals that have been used along this work.
 
\section{cMERA formalism}
\label{sec:formalism}

cMERA \cite{cMERA1, cMERA2} is a real space renormalization group procedure on the quantum state that builds a scale dependent wavefunctional $\Psi[\phi,u]$ in the Schr\"odinger picture given by,
\begin{equation}
\label{cMERAansatz}
\Psi[\phi,u]=\langle \phi|\Psi_u\rangle = \langle \phi|\, \mathcal{P} \, e^{-i \int_{u_{\text{IR}}}^u (K(u') + L) \, du' } \, |\Omega\rangle
\, ,
\end{equation}
where $u$ parametrizes the scale of the renormalization and $\mathcal{P}$ is the $u$-ordering operator.  Here $L$ is the dilatation operator and the generating operator $K(u)$ is the so-called ``entangler''. The renormalization scale parameter $u$ is usually taken to be in the interval $[u_{IR},u_{UV}] = (-\infty,0]$.  $u_{UV}$ is the scale at the UV cutoff $\epsilon$, and the corresponding momentum space UV cutoff is $\Lambda = 1/\epsilon$.  $u_{IR}$ is the scale in the IR limit.  

The state $|\Psi_{\Lambda}\rangle \equiv|\Psi_{ UV} \rangle$ is the state in the UV limit and it may be the ground state of a quantum field theory. The state $| \Omega \rangle$ is such that there is no entanglement between spatial regions upon which the cMERA flow builds correlations at successively smaller distance scales. We will impose $|\Omega \rangle$ to be scale invariant with respect to spatial dilatations, so that  $e^{-iLu}|\Omega \rangle=|\Omega \rangle$ or, equivalently,
\begin{align}
L |\Omega \rangle=0\, .
\end{align}
In this work, we will assume $|\Omega\rangle$ to satisfy the following condition
\footnote{Let us note that, traditionally, $\ket\Omega$ has been assumed a Gaussian state with no entanglement at the IR scale, $\ket{\Psi_{IR}}=\ket\Omega$ \cite{cMERA1}. Having a quadratic entangler that performs scale dependent Bogoliubov transformations on the fields of the free theory, the final state in the UV is also Gaussian. However, the treatment of interacting theories with truly non-Gaussian entanglers unveils a plethora of boundary conditions which may be physically more relevant. For instance, we can consider a reversed cMERA flow in which the pure Gaussian state is defined at the UV scale, $\ket{\Psi_{UV}}=\ket\Omega$ whereas the maximally entangled non-Gaussian state occurs at the IR. Precisely, such types of flows become essential for theories that exhibit asymptotic freedom. For example, the ground state of the Gross-Neveu model is expected to be Gaussian as the energy scale increases. We want to stress that this reversed and other interpolating flows can be smoothly implemented in our icMERA formalism once the scale of the Gaussian state has been fixed.}
\begin{align}
\label{scaleinv1}
\left(\sqrt{\omega_{\Lambda}} \,(\phi(\vp) - \chi_0 ) + \frac{i}{\sqrt{\omega_{\Lambda}}} \, \pi(\vp)\right) |\Omega\rangle = 0
\, ,
\end{align}
for all  momenta $\vp,$ where $\omega_{\Lambda}=\sqrt{\Lambda^2 + m^2}$ {with $m$ the mass of the particles in the free theory} and {$\chi_0=\langle \Psi_{\Lambda}|\phi(x)|\Psi_{\Lambda}\rangle$.} This state satisfies 
\begin{align}
\langle\Omega|\phi(\vp)\phi(\vq)|\Omega\rangle =\frac{1}{2\omega_{\Lambda}}\ \delta^{d}(\vp+\vq)\, ,
\quad
\langle\Omega|\pi(\vp)\pi(\vq)|\Omega\rangle=\frac{\omega_{\Lambda}}{2}\ \delta^d(\vp+\vq)\, .
\end{align}

The nonrelativistic dilatation operator $L$ does not depend on the scale $u$ and it is only governed by the scaling dimensions of the fields. It is taken as the ``free'' piece of the cMERA Hamiltonian and is given by
\begin{align}
L=-\frac{1}{2}\int\, d\vx \left[\pi(\vx)\left(\vx \cdot \nabla\phi(\vx)\right) + \left(\vx \cdot \nabla\phi(\vx)\right)\pi(\vx) + \frac{d}{2}\left(\phi(\vx)\pi(\vx) + \pi(\vx)\phi(\vx)\right)\right]\, .
\end{align}
The transformation properties of the field under the scale transformation $L$ are given by:
\begin{equation}
\begin{split}
e^{-iuL}\phi(\vp)e^{iuL}=&\ e^{-\f{d}{2}u}\phi(\vp e^{-u})\, , 
\\
e^{-iuL}\pi(\vp)e^{iuL}=&\ e^{-\f{d}{2}u}\pi(\vp e^{-u})\, .
\end{split}
\end{equation}

The entangler $K(u)$, which contains all the variational parameters to be optimized, creates entanglement between field modes with momenta $|\vp| < \Lambda$, where $\Lambda$ is a generic cutoff. Actually, as we will see in Section \ref{sec:i_cMERA_circuit}, various cutoff parameters can co-exist simultaneously, in such a way that they regulate the strength of each of the variational parameters at different regions of momentum space.
As a consequence, the entangler is considered the ``interacting'' part of the cMERA Hamiltonian. In our approach, only the choice of $K(u)$ will fully determine whether $\ket\Omega$ transforms into a Gaussian or a non-Gaussian state. For example,   entanglers containing only quadratic operators generate scale dependent Bogoliubov transformations on the field operators, and these transform Gaussian states into Gaussian states. Conversely, entanglers possessing higher order operators induce nonlinear transformations on the field operators which, when acting on Gaussian states, generate non-Gaussian states \cite{Cotler2,JJ1}.

The unitary operator in Eq. \eqref{cMERAansatz} 
\begin{equation}
U(u_1,u_2) \equiv \mathcal{P} \exp \left[ -i \int_{u_2}^{u_1} du \ (K(u) + L ) \right]\, 
\end{equation}
can be understood as a Hamiltonian evolution with $K(u)+ L$ along the scaling parameter $u$. As such, it is useful to define cMERA in the ``interaction picture'' through the unitary transformation of states
\begin{equation}
| \Phi_u \rangle  = e^{i L u} | \Psi_u\rangle\, .
\end{equation}
In this picture, the entangler is given by
\begin{equation}
\hat{K}(u) =  e^{i L \, u} \, K(u)  e^{-i L \, u}\, ,
\end{equation}
and the $u$-evolution is determined by the unitary operator
\begin{equation}\label{eq:cMERA_evol_free}
\cU(u_1,u_2) = \mathcal{P} \exp \left[ -i \int_{u_2}^{u_1} du \, \hat{K}(u) \right]\, .
\end{equation}
Thus, one may write the wavefunctional in the interaction picture as
\begin{equation}\label{cMERAansatz_ip}
\Phi[\phi,u]=\langle \phi|\, \cU(u,u_{IR}) | \Omega\rangle
= \langle \phi| \mathcal{P} e^{ -i \int_{u_{IR}}^{u} du' \, \hat{K}(u') } |\Omega \rangle \, .
\end{equation}

\subsection{Non-Gaussian States: Beyond Quadratic Entanglers}
\label{sec:bender-dunne}

As the ground states of interacting theories are generically non-Gaussian, our approach to construct cMERA for interacting theories consists of considering scale dependent non-Gaussian trial wavefunctionals. 
Such states will be generated by the action of an entangler that contains nonquadratic operators on a Gaussian state. Precisely, this formulation is in correspondence with the fact that in QFT, trial states created by introducing polynomial corrections to a Gaussian state represent a finite number of particles and those are suppressed in the thermodynamic limit. 
As a consequence, in going beyond the Gaussian ansatz, we are forced to use a class of variational extensive states for which the energy density does not depend on the volume. Regrettably, commutators of nonquadratic operators yield operators with increasingly larger products of $\phi$'s and $\pi$'s \cite{Cotler2} and hence, do not close as an algebra. This is a significant obstacle to systematically define unitaries that build non-Gaussian states from a Gaussian or another non-Gaussian state, which is precisely the crucial step to define cMERA flows for interacting field theories. 

To explain various proposals that aim to circumvent this problem, we will closely follow \cite{Cotler2}. Let us firstly consider a generic entangler, noting that any Hermitian operator $\cO$ in a scalar field theory can be written as
\begin{equation}
\label{eq:generator_op}
\cO = \sum_{n=0}^\infty \sum_{s=0}^n \int_{\vp_1 \cdots  \vp_n} \, c_{n}^{(s)}(\vp_1 \cdots  \vp_n) \, K_{n}^{(s)}(\vp_1 \cdots  \vp_n)
\end{equation}
where  $c_n^{(s)}$ are real-valued functions that will play the r\^ole of variational parameters, whereas  $K_{n}^{(s)}(\vp_1 \cdots  \vp_n)$ is defined as
\begin{align}
\label{eq:fieldbasis}
K_{n}^{(s)}(\vp_1 \cdots  \vp_n) \equiv \, \phi(\vp_1)\cdots \phi(\vp_s) \pi(\vp_{s+1}) \cdots \pi(\vp_n) 
 + \pi(\vp_{s+1}) \cdots \pi(\vp_n) \phi(\vp_1)\cdots \phi(\vp_s)\, .
\end{align}
Without loss of generality, we will translate this problem to (0+1)-dimensional quantum mechanics. Our conclusions can be straightforwardly translated to $(d+1)$-dimensional field theory. In this case, we introduce the Bender-Dunne algebra of Hermitian operators $\{T_{m,n}\}_{m,n=-\infty}^{+\infty}$, where \cite{Bender:1989yna,Bender:1989fs}
\begin{align}
T_{m,n}
\equiv
\frac{1}{2^n}\sum_{k=0}^\infty\frac{\Gamma(n+1)}{k!\ \Gamma(n-k-1)}x^k p^{m}x^{n-k}
=
\frac{1}{2^m}\sum_{j=0}^\infty\frac{\Gamma(m+1)}{j!\ \Gamma(m-j-1)}p^j x^n p^{m-j}
\ .
\end{align}
Let us consider the subalgebra $\{T_{m,n}\}_{m,n=0}^{+\infty}$ and study the unitaries of the form
\begin{align}
U=\exp\left\{
	i\sum_{m,n=0}^\infty c_{m,n} T_{m,n}
	\right\}
\ , \qquad
c_{m,n}\in\mathbb{R}
\ .
\end{align}
The order of an operator is defined as the largest value of $m+n$, for which $c_{m,n}\neq0$.
Motivated by cMERA, we are interested in calculating quantities like $\matrixel{\psi}{U^\dagger T_{m,n}U}{\psi}$, where $T_{m,n}$ would be any of the Hamiltonian terms. Then we will have
\begin{align}
\matrixel{\psi}{U^\dagger T_{m,n}U}{\psi}
=
\matrixel{\psi}{\exp\left(-i \ \text{ad}_{\sum_{p,q=0}^\infty c_{p,q} T_{p,q}}\right) T_{m,n} }{\psi}
\ ,
\end{align}
where we have used the Hadamard's lemma
\begin{align}
e^{X}Ye^{-X}=Y+\left[X,Y\right]+{\frac {1}{2!}}[X,[X,Y]]+{\frac {1}{3!}}[X,[X,[X,Y]]]+\cdots 
=e^{\operatorname {ad} _{X}}Y
\ .
\label{eq:hadamard}
\end{align}
Interestingly, one can show that the commutator of an order-2 and an order-$k$ operator gives rise to, at least, an order-$k$ operator:
\begin{align}
[\text{order }2,\ \text{order }k]\le \text{order } k
\ .
\label{eq:order-commutator}
\end{align}
In particular, for $k=2$, operators form the Lie algebra $\mathfrak{Heis}(3,\mathbb{R})\oplus \mathfrak{sl}(2,\mathbb{R})$, and generate unitaries that map Gaussian states to Gaussian states. Namely, if $Q$, $Q'$ are order $\le2$, then the commutator $[Q, Q']$ will also have order $\le2$. As a result, if $\ket\Psi$ is a Gaussian state, then $e^{-i Q} \ket\Psi$ is also a Gaussian state. In this sense, we note that the entangler operator in Gaussian cMERA \cite{MERA1} can be seen as the operator $K_2^{(1)}$ with a scale dependent variationally optimized coefficient $c_2^{(1)}(\vp_1,\vp_2; u)$. Actually, \eqref{eq:order-commutator} ensures that even interacting terms, as for example $\lambda\phi^4$, remain under control when transformed with unitaries containing order-2 entanglers, as it was done in \cite{Cotler1}. However, because trial states are still Gaussian, non-Gaussian effects will be absent. For example, quantities as kurtosis or connected 4-point correlators automatically vanish \cite{JJ1}.

When entanglers containing higher order operators are considered, the situation is less trivial: If the order of the entangler is higher than 2, then the quantity $U^\dagger T_{m,n}U$ is an operator of infinite order. Namely, according to \eqref{eq:order-commutator}, the nested commutators induce higher-order operators at every step which propagate out of control. This is why order-2 entanglers are the only ones that can be straightforwardly used on any free or interacting theory.

To get over this issue, various methods have been proposed. For example, a method based on the following unitary
\begin{align}
U
=
\exp\left\{
	-i\left(
		\sum_{m+n\le 2}c_{m,n} T_{m,n}
		+\epsilon \sum_{3\le p+q}c_{p,q} T_{p,q}
		\right)
	\right\}
\ ,
\label{unitary-perturbation}
\end{align}
where $\epsilon$ is assumed to be a small parameter, proves to be useful for a perturbative treatment. In \cite{Cotler2} it is shown how to perturbately obtain the ground state of the quantum mechanical toy model of the anharmonic oscillator $H=p^2/2m +mx^2/2 +\lambda x^4$ from a Gaussian state. In particular, upon Taylor expanding the unitary \eqref{unitary-perturbation} above $\epsilon=0$, identifying $\epsilon=\lambda$ and choosing a set of $T_{p,q}$ operators with $p+q>2$, the non-Gaussian ground state $\ket{\psi_{\text{anharm}}}$ is given by
\begin{align}
\ket{\psi_{\text{anharm}}}=U\ket{\psi_{\text{harm}}}+\cO(\lambda^2)
\ ,
\end{align}
where $\ket{\psi_{\text{harm}}}$ is the ground state of the harmonic oscillator. Based on these principles, several entanglers have been proposed in field theory at a perturbative level \cite{Cotler2,Bhattacharyya:2018bbv}.

In \cite{JJ1}, we have proposed a different approach to truncate the infinite series of higher order operators. Based on \cite{polley89,ritschel90}, we consider a cMERA entangler in terms of the following field theory anti-Hermitian operator:
\begin{align}
\cB(u)
=
	-s\int_{\vp \vq_1\cdots \vq_m} 
g(p,q_1,\ldots,q_m;u)\, \pi(\vp)\, \phi(\vq_1)\ldots\phi(\vq_m) \delta(\vp+\vq_1+\cdots\vq_m) 
\ ,
\end{align}
where $m\ge2$ is a parameter to be fixed.
\footnote{Despite the sum of various nonquadratic entanglers can be obviously considered, this is beyond the scope of this work.}
Here, $s$ is the variational parameter whereas $g(p,q_1,\ldots,q_m;u)$ consists of a combination of scale dependent variational cutoff parameters. The explicit dependence on the variational cutoff parameters will certainly be essential for the truncation of the series of nested commutators and for the optimization. In contrast to quadratic entanglers, this operator clearly goes beyond Bogoliubov transformations and induces nonlinear canonical transformations on the fields $\phi$ and $\pi$. It is important to emphasize that the variational parameter $s$ is not necessarily perturbative
\footnote{
Various nonperturbative effects have been obtained with finite $s$ \cite{polley89, ritschel90}}.
In addition to this, in Section \ref{sec:i_cMERA_circuit} we will show that, when applied, scale dependent non-Gaussian effects are unambiguously captured: the connected part of multi-point correlators, \emph{kurtosis} or \emph{skewness} clearly result scale dependent nonvanishing quantities \cite{JJ1}.

In summary, upon reviewing the cMERA formalism in full generality, we have reduced the problem of having (non-)Gaussian trial states to a particular choice of the operators entering the entangler. In the following sections we will review the Gaussian cMERA formalism and explain the construction of a scale dependent non-Gaussian cMERA circuit. Finally, we will perform the optimization on the $\lambda\phi^4$ theory and study various truly non-Gaussian observables.

\section{Gaussian cMERA}
\label{sec:gaussian}

\subsection{Quadratic Entangler}
\label{sec:quadratic}

For free scalar theories in $(d +1)$ dimensions with $d$, the spatial dimensions of the theory, $K(u)$ is given by the quadratic operator  \cite{cMERA1,cMERA2}
\begin{equation}
K(u) =
\frac{1}{2}\int_{\vp}\, g(p;u)\, \left[
\phi(\vp)\pi(-\vp)
+ \pi(\vp)\phi(-\vp)
\right]
\, ,
\label{eq:entangler}	
\end{equation}
{where $p\equiv|\vp|$ and $\int_{\mathbf{p}}\equiv\int\, (2\pi)^{-d}\, d^{d}p$.  The conjugate momentum of the field $\phi(\vp)$ is $\pi(\mathbf{p})\equiv-i\bar{\delta}/\delta\phi(-\vp)$, such that $[\phi(\mathbf{p}), \pi(\mathbf{q})] = i\bar{\delta}(\mathbf{p+q})\,$, with $\bar{\delta}(\mathbf{p})\equiv(2\pi)^{d}\delta(\mathbf{p})$. The function  $g(p;u)$ in (\ref{eq:entangler}) is the only variational parameter to be optimized in the cMERA circuit.} This function factorizes as 
\begin{equation}
\label{eq:free_var}
g(p;u)=g(u)\, \Gamma(p/\Lambda)\, ,
\end{equation} 
where  $\Gamma(x)\equiv \Theta(1-|x|)$ and $\Theta(x)$ is the Heaviside step function; $g(u)$ is a real-valued function known as \emph{density of entanglers} and $\Gamma(p/\Lambda)$ implements a high frequency cutoff such that $\int_{\vp} \equiv \int_{\vp}^{\Lambda}$ \cite{cMERA1, cMERA2}. The sharp cutoff function ensures that $K(u)$ acts locally in a region of size $\ep \simeq \Lambda^{-1}$. It is also possible to define the entangler $K(u)$ through localized smooth smearing functions instead of sharp cutoff functions \cite{magicmera}. 

In the interaction picture the entangler operator reads as,
\begin{align}
\hat{K}(u)
&=\f{1}{2}\int_{\vk}\, e^{du} \left[g(k;u)\phi(ke^u)\pi(-ke^u)+g(k,u)\pi(ke^u)\phi(-ke^u)\right]
\nonumber \\
&=\f{1}{2}\int_{\vk} \left[g(ke^{-u};u)\phi(k)\pi(-k)+g(ke^{-u},u)\pi(k)\phi(-k)\right]\, 
\ ,
\label{disein}
\end{align}
where the integral in the second line will be suppressed by the cutoff for $|\vk|\leq \Lambda e^{u}$.

With this, the operator $U_G(0,u)\equiv \mathcal{P}e^{ -i\int_{0}^{u} du' (K(u')+L)}$ defines the cMERA evolution in terms of the  scale-dependent linear transformation of the fields
\begin{align}
\label{eq:cMERA_fields}
U_G(0,u)^{-1} \phi(\vp)U_G(0,u)
=&\
e^{-f(p,u)}e^{-\frac{u}{2}d}\phi(\vp e^{-u})
\ ,
\\ \nonumber
U_G(0,u)^{-1} \pi(\vp)U_G(0,u)
=&\
e^{f(p,u)}e^{- \frac{u}{2}d}\pi(\vp e^{-u})\, ,
\end{align}
with $f(p,u) \equiv \int_0^u d u' \ g(p e^{- u'}; u')$. We have used the subscript $G$ because it is straightforward to show that the cMERA wavefunctional (\ref{cMERAansatz}) with the quadratic entangler (\ref{eq:entangler}) and a reference state such as (\ref{scaleinv1}) can be written as the Gaussian wavefunctional given by 
\begin{align}
\label{cMERAwavefunctional1shifted}
\Psi[\phi;u]_{SG} = 
N\, \exp \left(-\frac{1}{4}\int_{\vp} \, \left(\phi(\vp) -\chi_0\right) \, F^{-1}(p ; u) \, \left(\phi(-\vp) -\chi_0\right) \right)
\, ,
\end{align}
where $N$ is a normalization constant and the scale dependent Gaussian kernel $F(p; u)$ is defined through the variational cMERA parameter $g(p;u)$ by \cite{Cotler1}
\begin{align}
\label{dictionaryJMV}
F^{-1}(p;u) = 2\, \omega_\Lambda \, \exp \left(2 \int_0^u d u' \ g(p e^{-u'}, u')\right)\,.
\end{align}
We note that this wavefunctional is built as
\begin{equation}
\Psi[\phi;u]_{SG} = U_S(\chi_0)\, 
\underbrace{N\, \exp \left(-\frac{1}{4}\int_{\vp} \, \phi(\vp) \, F^{-1}(p ; u) \, \phi(-\vp) \right)}_{\Psi[\phi;u]_{G}}\, ,
\end{equation}
where the operator that shifts the argument of any functional (and specifically the Gaussian wavefunctional) by a constant $\chi_0$, is given by $U_S(\chi_0) = e^{\cO_S}$ with 
\begin{align}
\cO_S=-\int_{\vp}\chi_0\, \frac{\delta}{\delta\phi(-\vp)}\, .
\end{align} 
This is an standard type of transformation which leaves invariant the measure of integration in the functional path-integral formalism invariant, \emph{i.e.}, ${\rm D}\phi={\rm D}\phi'$ where $\phi'=\phi-\chi_0$, the shift of the variable of integration by a fixed background field configuration. A more general possibility, which will be analyzed in the next section, was introduced in \cite{polley89, ritschel90, ibanez} and amounts to shifting part of the field modes $\phi(\vk)$ by a nonlinear polynomial functional of other field modes.
In geometrical terms this is tantamount to  nonlinearly deform the infinite-dimensional configuration vectors $\lbrace |\phi(x)\rangle \rbrace$ of the field theory. For the case of free theories, generating the set of ``fixed-background'' shifted scale dependent cMERA Gaussian wavefunctionals \eqref{cMERAwavefunctional1shifted} is enough from a variational point of view.

\subsection{Free Scalar Theory}
\label{sec:free-gaussian}

In order to ``solve'' the cMERA ansatz for the relativistic free massive scalar theory, one applies the variational principle to minimize the energy functional given by
\begin{align}
H_{\Lambda}=\f{1}{2}\int_{\vk}  [\pi(\vk)\pi(-\vk)+\omega_k^2 \phi(\vk)\phi(-\vk)]\, , \label{haml}
\end{align}
with $\omega_k=\s{k^2+m^2}$ \cite{cMERA1, cMERA2}. The total energy $E$ is given by
 \begin{align}
 E
=&\ 
\langle \Psi_{\Lambda}|H_{\Lambda}|\Psi_{\Lambda} \rangle 
\nonumber
\\
=&\ \langle \Omega|H(u_{\mathrm{IR}})|\Omega\rangle 
 \nonumber \\
=&\ \langle \Omega| \int_{\vk}  \f{1}{2}
 \Big[e^{2f(k,u_{\mathrm{IR}})}e^{-u_{\mathrm{IR}}d}\pi(ke^{-u_{\mathrm{IR}}})\pi(-ke^{-u_{\mathrm{IR}}})
 \nonumber \\
 &\qquad \qquad +\omega_k^2
 e^{-2f(k,u_{\mathrm{IR}})}e^{-u_{\mathrm{IR}}d}\phi(ke^{-u_{\mathrm{IR}}})\phi(-ke^{-u_{\mathrm{IR}}}) \Big]|\Omega \rangle
 \nonumber \\
=&\ \int d^dx \int_{\vk} \f{1}{4}\left[e^{2f(k,u_{\mathrm{IR}})}\omega_{\Lambda}+\f{\omega_k^2}{\omega_{\Lambda}}e^{-2f(k,u_{\mathrm{IR}})}\right]\ .
 \end{align}
 Then, we impose
 \begin{align}
 \f{\delta E}{\delta g(u)}=\int d^d x\int_{\vk} \left( e^{2f(k,u_{\mathrm{IR}})}\omega_{\Lambda}-\f{\omega_k^2}{\omega_{\Lambda}}e^{-2f(k,u_{\mathrm{IR}})}\right)
 \Gamma(ke^{-u}/\Lambda)=0\, ,
 \end{align}
 which yields
 \begin{align}
 f(k,u_{\mathrm{IR}})=\f{1}{2}\log \f{\omega_k}{\omega_{\Lambda}}, \qquad
 k<\Lambda \ . 
 \end{align}
 Using that
 \begin{align}
 f(k,u_{\mathrm{IR}})=\int^{u_{\mathrm{IR}}}_0 g(ke^{-u};u)\, du=\int^{-\log \Lambda/k}_0 g(u)\, du \ ,
 \end{align}
one obtains
 \begin{align}
  g(u)=\f{1}{2}\,\left(\f{k\, \partial_{k}
 	\omega_k}{\omega_k}\right)\Biggr|_{k=\Lambda e^u} = \f{1}{2} \f{\Lambda^2 e^{2u}}{(\Lambda^2 e^{2u}+m^2)}\, , \label{g_variational} 
 \end{align}
and
\begin{align}
f(k,u)
=
\left\{
\begin{array}{llll}
\displaystyle
\f{1}{4}\log \f{\Lambda^2e^{2u} + m^2}{\omega_{\Lambda}^2}&, \ \ \ & k\leq\Lambda e^{u} &,
\nonumber \\
\displaystyle
\f{1}{4}\log \f{k^2+m^2}{\omega_{\Lambda}^2} &, \ \ \  & k>\Lambda e^{u}& .
\end{array}
\right.
\end{align}

\subsection{Self-Interacting Scalar $\lambda\phi^4$ Theory}
\label{sec:interacting-gaussian}
As commented in Section \ref{sec:Intro} and based on the discussion of Section \ref{sec:bender-dunne}, the Gaussian ansatz has been widely used in the context of interacting field theories. In this sense, in \cite{Cotler1}, a cMERA circuit based on the quadratic entangler \eqref{eq:entangler} was used to study the self-interacting scalar theory. This model has a mass gap and flows to a free theory in the IR, where the IR ground state is exactly a Gaussian wavefunctional.  Similar to the free case, we minimize the expectation value of the Hamiltonian 
\begin{align}
H_{\Lambda} = \int d^d x \, \left[\frac{1}{2} \pi^2 + \frac{1}{2}(\nabla \phi)^2 + \frac{m^2}{2} \, \phi^2 + \frac{\lambda}{4!} \, \phi^4 \right]\, , 
\end{align}
with respect to the ansatz wavefunctionals $\Psi[\phi,u]$ of the form \eqref{cMERAwavefunctional1shifted}. As shown in \cite{Cotler1}, to solve the variational problem we compute
\begin{align}
E&=\langle \Psi_{\Lambda}|H_{\Lambda}|\Psi_{\Lambda} \rangle \\ \nonumber
& = \frac14 \int_\vp \Big[
F(p)^{-1}
+p^2\, F(p)
\Big]
+\frac12m^2 (I+\chi_0^2)
+\frac{\lambda}{4!}\left(
\chi_0^4
+6I\chi_0^2
+3I^2
\right)\, ,
\end{align}
with $I = \frac12 \int_{\vp}\, F(p)$. Then we impose $\delta E/\delta F(p)=0$, which yields an optimized Gaussian kernel $F(p)$,
\begin{align}
\label{eq:cMERA_JJM}
 F(p) &= \frac{1}{\sqrt{p^2 +  \mu^2}}
\, ,
\end{align}
where $\mu$ is the modified mass of the propagating Gaussian quasi-particles
\begin{equation}
\label{gapeqn1}
\mu^2 = m^2 +\frac{\lambda}{2}\left(\chi_0^2+I\right)=m^2 + \frac{\lambda}{2}\left(\chi_0^2 + \int_{\vp}\, \frac{1}{2 \sqrt{p^2 + \mu^2}} \right)\,.
\end{equation}
Finally, through \eqref{dictionaryJMV}, we obtain the optimized Gaussian cMERA variational parameter
\begin{align}
\label{gsol1}
g(p;u) =& \, \frac{1}{2} \, \frac{e^{2u}}{e^{2u} + \mu^2/\Lambda^2} \,  \Gamma(p/\Lambda)\, .
\end{align}

Remarkably, the wavefunctional optimization over an infinite-dimensional space of kernels $F$ has reduced to solving a single nonlinear equation for $\mu$.  The Gaussian cMERA wavefunctional thus obtained is a vacuum state for a free theory with mass given by \eqref{gapeqn1}.   The optimized ansatz captures all 1-loop $2$-point correlation functions, and additionally captures the resummation of all cactus-like diagrams \cite{GEP1}. Nevertheless, the ansatz is unable to capture the effects of the interaction on higher order correlation functions. Let us ellaborate on this point.  

Quantum field theories are characterized by correlation functions of the form
 \begin{align}
 G^{(N)}(\vx_1,\cdots,\vx_N)=\expval{\cO(\vx_1)\cdots\cO(\vx_N)}
 \, ,
 \end{align}
 where $\cO$ is a field operator, $N$ is the order of the correlation and, in general, the expectation value is taken with respect to the ground state of the system. In absence of interactions, the relevant information is encoded only in the 2-point correlation function $G^{(2)}$, with higher-order correlations $G^{(N>2)}$ decomposing as a sum of products of only $G^{(N\leq 2)}$. This is the case when the ground state of the system is Gaussian. In order to characterize the influence of interactions, it is useful to  decompose the $N$-th point correlator $G^{(N)}$ into
 \begin{align}
 G^{(N)}(\vx_1,\cdots,\vx_N)=G^{(N)}_{\rm dis}(\vx_1,\cdots,\vx_N) + G^{(N)}_{\rm con}(\vx_1,\cdots,\vx_N)\, .
 \end{align}
 $G^{(N)}_{\rm dis}$ refers to the disconnected part of the correlation function and it is determined by lower order correlation functions $G^{(N'<N)}$. In this sense, the $G^{(N)}_{\rm dis}$ does not possess any new information of the system at order $N$. On the other hand, $G^{(N)}_{\rm con}$, the connected part of the correlation function, has access to proper and characteristic information about the system at order $N$. As a result, a complete  factorization of higher-order correlation functions as in the Gaussian case, is equivalent to have $ G^{(N\geq2)}_{\rm con}=0$. In other words, the connected part of $N$-point  correlation functions distinguishes the ground  states of interacting theories from those of noninteracting ones: \emph{i.e.}, while for Gaussian states the connected correlation functions of order higher than two vanish, those of interacting systems are generally nonzero. This is the reason for which, despite being successful in capturing some nonperturbative effects through the gap equation, it would desirable, in order to study the phenomena of strongly interacting systems, to have powerful tools that are able to quantify the effect of perturbative and especially, nonperturbative quantum processes that contribute to the connected part of $N$th-point correlators.

\section{Non-Gaussian cMERA}
\label{sec:ng-cMERA}
In this section we present a systematic and model independent
\footnote{
While this method always generates a set of non-Gaussian trial states irrespective of the model, picking a particular theory will only be relevant for determining the optimal values of the variational parameters. For example, see \cite{JJ1} for theories with nonpolynomial potentials.
}
 formalism to go beyond the Gaussian approach in interacting field theories. This formalism  was generalized to nonperturbatively build non-Gaussian cMERA wavefunctionals in \cite{JJ1}. The method relies in performing a set of nonlinear transformations on the fields of the theory that extends the linear transformations defining the Gaussian approach for free theories. 

\subsection{Nonlinear Canonical Transformations}
\label{sec:nlct}

According to \cite{polley89,ritschel90,ibanez}, extensive non-Gaussian trial states wavefunctionals can be built as
\begin{align}\label{eq:NG_wavefunc}
\tilde \Psi[\phi] =\tilde U\, \Psi_G[\phi]=\exp(\cB)\, \Psi_G[\phi] \, ,
\end{align}
where $\Psi_G[\phi]$ is a normalized Gaussian wavefunctional and $\tilde U=\exp(\cB)$, with $\cB^{\dagger} = -\cB$ an anti-Hermitian operator that nonperturbatively adds new variational parameters to those in the Gaussian wavefunctional. As it will be shown below, the expectation value of any operator $\mathcal{O}(\phi,\pi)$ with these states amounts to the calculation of a Gaussian expectation value for the transformed operator $\widetilde{\mathcal{O}}=\tilde U^{\dagger}\, \mathcal{O}\, \tilde U$. Remarkably, a suitable choice of $\cB$, while  leading  to a non-Gaussian trial state, can indeed truncate the commutator expansion in Hadamard's lemma\footnote{${\rm Ad}_{\cB}\, (A)\equiv \exp(\cB)\, A\, \exp(-\cB)$,  ${\rm ad}_{\cB}(A)\equiv [\cB,A]$.}, 
\begin{align}
\widetilde{\mathcal{O}}= {\rm Ad}_{\cB}\, (\mathcal{O})= e^{{\rm ad}_{\cB}}\, \mathcal{O}\, .
\label{eq:conm_series}
\end{align}
This reduces the calculation of expectation values of functionals to a finite number of Gaussian expectation values. The exponential nature of $\tilde U$ ensures the correct extensive volume dependence of observables and specifically the total energy of the system. Furthermore, as $\tilde U$ is unitary, the normalization of the state is preserved. The operator $\cB$ consists of a product of $\pi$'s and $\phi$'s, which is given by
\begin{align}\label{eq:BBos}
\mathcal{B}
= 
-s\int_{\vp \vq_1\cdots \vq_m} 
h(p,q_1,\ldots,q_m)\, \pi(\vp)\, \phi(\vq_1)\ldots\phi(\vq_m) \delta(\vp+\vq_1+\cdots\vq_m) 
\ ,
\end{align}
with $m \in \mathbb{N}$. We will denote these operators from here in advance symbolically  as $\mathcal{B}\equiv \pi\, \phi^{m}$. Here, $s$ is a  variational parameter that, as it will be shown later, tracks the deviation of any observable from the Gaussian case. $h(p,q_1,\ldots,q_m)$ is a variational function that must be optimized upon energy minimization.
It is symmetric w.r.t. exchange of $q_i$'s  and is constrained to satisfy:
\begin{align}
\label{eq:constraint}
h(p,q_1,\ldots,q_m)
=0 
\ , 
\quad p=q_i\, ,
\qquad\text{and} \qquad
h(p,q_1,\ldots,q_m)\, h(q_i,k_1,\ldots,k_m) =&\ 0
\ .
\end{align}
These conditions ensure that the commutator series terminates after the first nontrivial terms. Namely, the constraints in  \eqref{eq:constraint} are the responsible for this truncation when the Hadamard's lemma \eqref{eq:hadamard} is applied.
The action of $\tilde U$ on the canonical field operators $\phi(\vp)$ and $\pi(\vp)$ is given by
\begin{eqnarray}
\label{trasfields}
	\tilde \phi(\vp)&\equiv&\tilde U^\dagger\, \phi(\vp)\, \tilde U 
	=
	\phi(\vp)
	+s \, \bar{\phi}(\vp)
	\ ,
	\\ \nonumber
	\tilde \pi(\vp)&\equiv&\tilde U^\dagger\, \pi(\vp)\, \tilde U
	=
\pi(\vp)
	-\, s \, \bar{\pi}(\vp)
	\ ,
\end{eqnarray}
where the quantities with a bar are defined as the nonlinear field functionals, 
\begin{equation}
\begin{aligned}
\bar{\phi}(\vp)\equiv&\ \int_{\vq_1\cdots \vq_m} h(p,q_1,\ldots,q_m)\, \phi(\vq_1)\cdots \phi(\vq_m) \delta(\vp-\vq_1-\cdots\vq_m) \ , 
\\ 
\bar{\pi}(\vp)
	\equiv&\
	m\, \int_{\vq_1\cdots \vq_m}
	h(q_1,p,\ldots,q_m)\, \pi(\vq_1)\, \phi(\vq_2)\phi(\vq_m) 
	\delta(\vp-\vq_1-\cdots\vq_m) 
\ .
\end{aligned}
\label{detail_transfields}
\end{equation}
Thus, we are considering a class of field transformations, 
\begin{align}
\tilde \phi \equiv \mathcal{F}(\phi;\Lambda)\, ,
\end{align}
dependent on variational parameters $s$ and $h$, which amount to shifting part of the field modes of $\phi$ by a nonlinear polynomial functional (degree $m$) of other field modes. The function $\mathcal{F}$ also depends on the energy scale.  As it will be shown later, the method of nonlinear transformations essentially provides a nonperturbative expansion of the physical observables of the theory under consideration, about a point-like Gaussian free field theory. 

Being $\tilde U$ unitary, the canonical commutation relations (CCR) still hold under the nonlinear transformation of the fields \eqref{trasfields} and \eqref{detail_transfields} giving,
\begin{align}
 [\tilde \phi(\vp), \tilde \pi(\vq)]=i\bar{\delta}(\vp + \vq)\, .
\end{align}

The constraints \eqref{eq:constraint} on the non-Gaussian variational parameter $h(p,q_1,\ldots,q_m)$ can be accomplished by taking the decomposition 
\begin{align}
h(p,q_1,\ldots,q_m) = \eta(p)\cdot \zeta(q_1) \cdot  \zeta(q_2)\cdots  \zeta(q_m)\, ,
\end{align}
where it is imposed that $\eta(p)\cdot \zeta(p) = 0$, \emph{i.e.}, the domains of momenta, where $\eta$ and $\zeta$ are different from zero have to be disjoint, up to sets of measure
zero. A suitable ansatz for $\eta$ and $\zeta$ is given by \cite{polley89, ritschel90}
\begin{eqnarray}
\label{eq:cutoff_ansatz}
\eta(p)&=&\Gamma((p/\Delta_1)^2)\, , \\ \nonumber
\zeta(q_i) &=& \left[\Gamma((\Delta_1/q_i)^2)-\Gamma((\Delta_2/q_i)^2)\right]\, ,
\end{eqnarray}
where  $\Delta_i$ ($i=1,2$), are variationally optimized, coupling dependent momentum cutoffs such that $0<\Delta_{i}\le\Lambda$. $\Gamma(x)$ refers to the sharp cutoff function in \eqref{eq:free_var}.

Summarizing, the method of nonlinear canonical transformations builds variational non-Gaussian trial wavefunctionals by applying the operator $\tilde U=\exp\left(\mathcal{B}\right)$ defined through the variational function $h(p,q_1,\ldots,q_m)$ to a Gaussian wavefunctional with a variational kernel $F(\vp)$. Being a model independent formalism, the explicit dependence of these  parameters on the interaction couplings of a theory is established through energy minimization. This will be discussed later for some concrete examples. 

\subsubsection*{Wavefunctionals: the generality of the ansatz}
We would like to analyze the effect of the transformation $\tilde U$ on wavefunctionals. These amount to probability amplitudes for concrete field configurations. For the ground state of a theory, the Hamiltonian completely determines the wavefunctional and, as ever, two competing and opposite trends act on configuring it. First, the kinematic term of the Hamiltonian favors a soft amplitude where many distinct field configurations have high probability. Contrarily, the potential term leverages a strong localization of the field configurations around the classical ground state. The tradeoff results in a state with maximum probability at the classical solution and decreasing amplitude for other configurations. For Gaussian functionals such as,
\begin{align}
\Psi_{G}[\phi]
=
N \exp\left(
-\f{1}{4} \int_\vk  \phi(\vk)  F^{-1}(\vk) \phi(-\vk)
\right)
\ ,
\end{align}
the half-mean width of the functional is proportional to $(k^2+\mu^2)^{-1/4}$. Thus, as the variational mass $\mu$ increases in the Gaussian kernel $F(\vk)$,  a stronger suppression for nonclassical configurations than in the free case occurs. This is what happens when the Gaussian ansatz is used, for instance, in the $\lambda \phi^4$ theory \cite{Cotler1}. That said, we now illustrate the action of $\tilde U$ on a Gaussian wavefunctional by choosing the transformation $\mathcal{B}=\pi\, \phi^2$ for clarity:
\begin{equation}
\begin{aligned}
	\tilde \Psi[\Phi] 
\equiv&\ 
\tilde U\, \Psi_G[\phi] = \tilde U\, \left(1 - \f{1}{4} \int_\vk \phi(\vk) F^{-1}(\vk) \phi(-\vk) +\cdots\right)
\\
=&
\ 1-	\f{1}{4} \int_\vk \left(	\phi(\vk)	-s\int_{\vq_1 \vq_2} h(k,q_1,q_2)\phi(\vq_1)\phi(\vq_2) \delta(\vk-\vq_1-\vq_2)	\right)\times 
\\ 
&\
\times  F^{-1}(\vk)\left(	\phi(-\vk)	-s\int_{\vq_3 \vq_4} h(k,q_3,q_4)\phi(\vq_3)\phi(\vq_4) \delta(-\vk-\vq_3-\vq_4)\right)+\cdots
\\ 
=&\
1-\f{1}{4} \int_\vk \Phi(\vk)  F^{-1}(\vk) \Phi(-\vk) +\cdots
\\
=&\
\Psi_G\left[\Phi(\vk)\right]
	\ ,
\end{aligned}
\end{equation}
where ellipses stand for the expansion of the exponential, $\Phi(\vk)\equiv \phi(\vk) -s\, \bar{\phi}(\vk)$ and $\bar{\phi}(\vk)$ corresponds to \eqref{detail_transfields} for $m=2$.
The result $\tilde U\, \Psi_G[\phi]=\Psi_G\left[\Phi(\vk)\right]$ shows that one is left with an \emph{effective} trial Gaussian state $\Psi_{G}$ that involves completely different fields $(\Phi)$ than the underlying \emph{microscopic} elementary fields which define the short distance dynamics of the theory $(\phi)$. To be more precise, the transformation $\tilde U$ generates a translation of the argument in the configuration space of the theory that symbolically reads as $\tilde \Psi[\phi]=\Psi_G[\phi-s \phi^2]$. 

Hence, while for the Gaussian case the decaying slope of the wavefunctional is
\begin{align}
 \log \Psi_{G}[\phi]=-\f{1}{4}\, \int_{\vk} F^{-1}(\vk)\, |\phi(\vk)|^2\, ,
\end{align}  
now the  decaying slope is corrected by new terms as
\begin{align}
\log \tilde \Psi[\Phi]= \log \Psi_{G}[\phi] +\f{s}{2} \, \int_{\vk} F^{-1}(\vk)\, \phi(\vk) \bar \phi(\vk) -\f{s^2}{4}\, \int_{\vk} F^{-1}(\vk)\, |\bar \phi(\vk)|^2 \, .
\end{align}  

\subsubsection*{Observables: the calculability of the ansatz} 

Now we focus on the action of $\tilde U$ on observables. The expectation value of any operator $\mathcal{O}(\pi, \phi)$ reduces to the calculation of a Gaussian expectation value for the transformed operator $\widetilde{\mathcal{O}}=\tilde U^{\dagger}\, \mathcal{O}\, \tilde U$, as we have
\begin{align}
\langle \tilde \Psi |\mathcal{O} | \tilde \Psi\rangle = \langle \Psi_{G} |\tilde U^{\dagger}\, \mathcal{O}\, \tilde U | \Psi_{G}\rangle\, .
\end{align} 
In this sense, it is of particular interest to consider $n$-point correlation functions. In general, we will have
\begin{align}
\expval{\phi(\vx_1)\cdots\phi(\vx_n)}_{NG}
=
\int_{\vp_1\cdots\vp_n}
e^{i\sum_k \vp_k \vx_k} \expval{\phi(\vp_1)\cdots\phi(\vp_n)}_{NG}
\ .
\end{align}
where the subscript $NG$ refers to an expectation value taken w.r.t. the non-Gaussian state \eqref{eq:NG_wavefunc}. To evaluate this, we use \eqref{trasfields} and \eqref{detail_transfields}, which yields
\begin{dmath}
\expval{\phi(\vp_1)\cdots\phi(\vp_n)}_{NG}
	=
	\expval{\tilde U^\dagger\, \phi(\vp_1)\, \tilde U \cdots \tilde U^\dagger\, \phi(\vp_n)\, \tilde U}_{G}
	=\ \expval{\phi(\vp_1)\cdots \phi(\vp_n)}_{G}\\ \nonumber
	+ s\left [\expval{\bar{\phi}(\vp_1)\phi(\vp_2)\cdots \phi(\vp_n)}_{G} +\cdots +\expval{\phi(\vp_1)\cdots \phi(\vp_{n-1})\bar{\phi}(\vp_n)}_{G}	\right]
	+s^2\left[ \expval{\bar{\phi}(\vp_1)\bar{\phi}(\vp_2)\phi(\vp_3)\cdots \phi(\vp_n)}_{G}
	+\cdots +\expval{\phi(\vp_1)\cdots\bar{\phi}(\vp_{n-1})\bar{\phi}(\vp_{n})}_{G}\right]
\\
\mbox{} \ \  \, \, \vdots
\\
	+ s^n \expval{\bar{\phi}(\vp_1)\cdots\bar{\phi}(\vp_n)}_{G}
	\ .
\end{dmath}
That is to say, the calculability of the ansatz allows to compute the expectation value of observables such as correlation functions in terms of a finite number of Gaussian expectation values. In particular, the terms proportional to $s^j$ in the non-Gaussian $n$-point correlation function correspond to $(n+m(j-1))$-point  Gaussian correlators, where $j=0,\ldots,n$ and $m$ is the power associated to the operator $\cB=\pi\phi^m$.

\subsection{Non-Gaussian cMERA Formalism}
Following \cite{JJ1}, now we use the method depicted above to generate cMERA non-Gaussian trial states $\tilde \Psi[\phi, u]$ implementing a renormalization group flow of the wavefunction for interacting field theories. This can be cast as
\begin{align}
\tilde \Psi[\phi, u]=\langle \phi |\tilde \Psi_u\rangle =  \matrixel{ \phi}{ \mathbb{U}(u)}{ \Omega} 
\ ,
\qquad\qquad \mathbb{U}(u) \equiv \tilde U\,  U_G(u) 
\ ,
\end{align}
where $ U_G(u)=U_G(u,u_{IR}) = \mathcal{P} \, e^{-i \int_{u_{\text{IR}}}^u (K(u') + L) \, du' }$ is the cMERA unitary operator for the free theory defined through the variational parameter \eqref{eq:free_var}
and $\tilde U=\exp\left(\mathcal{B}\right)$. To go beyond Gaussian approach, we have to consider operators $\mathcal{B}$ that at least are cubic in the products of $\pi$ and $\phi$ fields. Here, we are focusing on the simplest case $\mathcal{B}=\pi\phi^2$. Recalling \eqref{eq:cMERA_fields},  the transformation $\mathbb{U}$ acts on the fields as follows:
\begin{equation}
\label{eq:nlct_u}
\begin{aligned}
	\mathbb{U}(u)^\dagger \phi(\vp)\, 	\mathbb{U}(u)
	=&\
	e^{-f(p,u)}e^{-\frac d2 u}\left[
	\phi(e^{-u}\vp)
	+s e^{\frac d2 u} \int_{\vq_1\vq_2} \bar{h}(p, q_1 , q_2;u)\, \phi(e^{-u}\vq_1)\phi(e^{-u}\vq_2) 	
	\right]  \, ,
	\\[10pt] 
	\mathbb{U}(u)^\dagger \pi(\vp)\, 	\mathbb{U}(u)
	=&\
	e^{+f(p,u)}e^{-\frac d2 u}\left[
	\pi(e^{-u}\vp)
	-2\, s e^{\frac d2 u} \int_{\vq_1\vq_2} \bar{h}(q_1, p , q_2;u)\, \pi(e^{-u}\vq_1)\phi(e^{-u}\vq_2) 	
	\right] 
   \, ,
\end{aligned}
\end{equation}
where, for compactness, we have dropped $\delta(\vp-\vq_1-\vq_2)$ appearing inside the integrals and we have defined
\begin{align}
\bar{h}(p, q_1 , q_2;u)
\equiv
h(p,q_1,q_2)\, e^{f(p,u)-f(q_1,u)-f(q_2,u)}
\ .
\label{eq:non-Gaussian-variational}
\end{align}
From a cMERA point of view, $\bar{h}(p,q_1,q_2; u)$ can be interpreted as a variational, scale and coupling dependent momentum cutoff function. While this happens automatically in the nonperturbative cMERA circuit, let us elaborate more on this now. In the Gaussian entangler the variational parameter $g(p;u)$ was decomposed in \eqref{eq:free_var} as 
\begin{align}
g(p;u)=g(u)\cdot \Gamma(p/\Lambda)\, ,
\end{align}
\emph{i.e.}, a scale dependent function $g(u)$ times a sharp momentum cutoff function $\Gamma(p/\Lambda)$. Similarly, we would expect the nonlinear transformations on the field modes \eqref{eq:nlct_u} (and thus, any observable built from products of these field modes) to have a similar structure,
\begin{eqnarray}
\bar{h}(p,q,r; u)
&\equiv & 
g(p,q,r;u)\cdot \Gamma_{\cB} (p,q_1,q_2)
\ ,
\end{eqnarray}
where $\Gamma_{\cB}$ is a generalization of the $\Gamma$ cutoff function in \eqref{eq:free_var}.
However, from \eqref{eq:non-Gaussian-variational} and \eqref{eq:cutoff_ansatz} we identify
\begin{equation}
\begin{aligned}
g(p,q,r;u)
=&\
e^{f(p,u)-f(q,u)-f(r,u)}
\ ,
\\
\Gamma_{\cB} (p,q,r)
=&\
h(p,q,r)=\eta(p)\cdot \zeta(q)\cdot \zeta(r)
\ .
\end{aligned}
\end{equation}
Then, $\bar{h}(p,q,r; u)$, which is a variational coupling dependent momentum cutoff function, implies the optimization of both the Gaussian parameter $f(p;u)$ and the cutoff momenta $\Lambda_{0,i}$. As we will see in Section \ref{sec:i_cMERA_circuit}, this variational scheme captures nonperturbative and non-Gaussian interaction effects, which turn out to be essential at the regime at which the Gaussian quasi-particle picture is no longer valid.
\footnote{
The construction of the non-Gaussian cMERA circuit, which is slightly more general, will be explained in Section \ref{sec:i_cMERA_circuit}.}

With this, in \cite{JJ1}, using the set of scale dependent nonlinear transformations given in \eqref{eq:nlct_u}, nonperturbative cMERA states for interacting field theories were built where, as before, it is straightforward to show that the non-Gaussian scale-dependent wavefunctional $\tilde \Psi[\phi;u]$ can be written as 
\begin{align}
 \tilde \Psi[\phi;u]= \tilde U\, \Psi_{G}[\phi;u] 
 =
 \Psi_G\left[\tilde \phi(\vp,u)\right] \, ,
\end{align}
with 
\begin{align}
 \tilde \phi(\vp,u)
=
 	e^{-f(p,u)}e^{-\frac d2 u}\left[
 \phi(e^{-u}\vp)
 -s e^{\frac d2 u} \int_{\vq_1\vq_2} \bar{h}(p, q_1 , q_2;u)\, \phi(e^{-u}\vq_1)\phi(e^{-u}\vq_2) 	
 \right]  
 \ .
\end{align}
Nevertheless, from a circuit/tensor network viewpoint, this methodology seems incomplete. Namely, the sequence of scale dependent non-Gaussian wavefunctionals $\tilde \Psi[\phi; u]$ is not generated by a cMERA circuit with an entangler containing the whole set of transformations, including the  nonquadratic ones. This would be a satisfactory and more natural continuum generalization of the lattice MERA algorithms. However, this is trivially guaranteed, as the product of two unitaries turns out to be another unitary. In this sense, an important result of this work is to explicitly obtain such unitary operator using the firm ground provided by the conceptual framework presented in this section.

\subsection{Fermionic non-Gaussian cMERA}
\label{sec:cMERA_fermions}

Before going into the proposal of a cMERA circuit that implements the scale dependent nonlinear canonical transformations for an interacting scalar field, let us comment on a formulation of nonlinear canonical transformations for fermionic fields. As advanced in \cite{JJ1}, the method to generate non-Gaussian cMERA wavefunctionals applies to fermionic field theories. Here we present some transformations.

Despite they can be generalized to any dimension and/or fermion, in this work, we will consider non-Gaussian transformations for two-dimensional Dirac fermions. Being model independent, our proposal is specially well suited to analyze the Gross-Neveu model (GN)\cite{grossneveu}. This  is a renormalizable, asymptotically free two-dimensional theory which displays chiral symmetry breaking and dynamical mass generation. The model describes $N$ flavors of massless spin 1/2 fermions in one spatial dimension with an attractive short range potential. Fermions get bound by the attractive force, and the fermion pair composite condenses and breaks a discrete $\mathbb{Z}_2$ symmetry. Because of the condensate, fermions dynamically acquire a mass. Indeed, the dynamics resembles the same mechanism as it occurs in four-dimensional QCD or BCS theory of superconductivity.
In Euclidean spacetime, the model with $N=1$ flavors of fermions is defined by the action
\begin{align}
S\left[\bar\psi,\, \psi\right]=\int d^2x\, \left[\bar\psi(x)\gamma^\mu\partial_\mu \psi(x) -\frac{g_0}{2}\left(\bar\psi(x) \psi(x)\right)^2\right]\, ,
\end{align}
where $\psi$ and $\bar \psi\equiv\psi^\dagger\gamma^0$ are two-dimensional Dirac spinor fields, $g_0$ denotes the coupling constant and the $\gamma$-matrices are given by
\begin{align}
\gamma^0=\sigma^1
\ , 
\qquad
\gamma^1=i\sigma^2
\ ,
\qquad
\gamma^3=\gamma^0\gamma^1
=\sigma^3
\ ,
\end{align}
with $\sigma^ i$ being the Pauli matrices. It is useful in order to deal with the GN model, to decompose the Dirac spinor  $\psi$, into its chiral projections, $\psi=\psi_L+\psi_R$, 
\begin{align}
\psi_L\equiv P_L\psi=\frac12(1-\gamma_3)\psi
\ , 
\qquad
\psi_R\equiv P_R\psi=\frac12(1+\gamma_3)\psi
\ .
\end{align}
In terms of the chiral projections, the GN model action can be written as
\begin{align}
S\left[\psi_L,\, \psi_R\right]=\int d^2x\, \left[\bar\psi_L\gamma^\mu\partial_\mu \psi_L + \bar\psi_R\gamma^\mu\partial_\mu \psi_R -\frac{g_0}{2}\left(\bar\psi_R \psi_L\right)^2-\frac{g_0}{2}\left(\bar\psi_L \psi_R\right)^2\right]\, .
\end{align}

As in the bosonic case, we build extensive non-Gaussian fermionic trial states as wavefunctionals of the form
\begin{align}\label{eq:NG_ferm_wavefunc}
\tilde \Psi[\psi_L,\psi_R] =\tilde U_F\, \Psi_G[\psi_L,\psi_R]=e^{\cF}\,  \Psi_G[\psi_L,\psi_R] \, ,
\end{align}
where $ \Psi_G[\psi_L,\psi_R]$ is a normalized Gaussian wavefunctional and $\tilde U_F=\exp(\cF)$, with $\cF$, an Hermitian operator that nonperturbatively adds the new variational parameters $s$ and $g(\vq_1,\cdots,\vq_4)$ to those in the Gaussian wavefunctional. Here, we consider the transformation 
\begin{eqnarray}\label{eq:fermionic_transform}
	\cF
	&=&s\, \int_{\vq_1\cdots\vq_4}g(\vq_1,\cdots,\vq_4)\bar\psi_R(\vq_1)\gamma^0\psi_L(\vq_2)\bar\psi_R(\vq_3)\gamma^0\psi_L(\vq_4)\delta(\tsum_i\vq_i)
	\ .
\end{eqnarray}
Using the Fierz identities, we can prove that momenta $\vq_2$ and $\vq_4$ can be exchanged by including an extra minus \cite{Ortin:2004ms,zinn_justin_book}. From here in advance we will drop the momentum conservation delta. The Hermiticity of $\cF$ imposes
\begin{align}
g(\vq_1,\vq_2,\vq_3,\vq_4)=g^*(\vq_2,\vq_1,\vq_4,\vq_3)
\ .
\end{align}

Upon this constraint, the effect of the transformation on the chiral spinor fields is given by:
\begin{align}
\Psi_{L,R}(\vp)\equiv \tilde U_F^\dagger \ \psi_{L,R}(\vp)\ \tilde U_F
\ ,
\end{align}
where, for the spinor index $\alpha$, we have
\begin{equation}
\begin{array}{rcll}
\Psi_{L\alpha}(\vp)
&=&
\psi_{L\alpha}(\vp)
-2 s\, \int_{\vq_1\vq_2\vq_3} g(\vq_1,\vq_2,\vp,\vq_3)\left[\bar\psi_R(\vq_1)\gamma^0\psi_L(\vq_2)\right]\psi_{L\alpha}(\vq_3)
& ,
\\[5pt]
\Psi_{R\alpha}(\vp)
&=&
\psi_{R\alpha}(\vp)
& ,
\\[5pt]
\bar\Psi_L^\alpha(\vp)
&=&
\bar\psi_L^\alpha(\vp)
& ,
\\[5pt]
\bar\Psi_{R}^{\alpha}(\vp)
&=&
\bar\psi_{R}^{\alpha}(\vp)
+2 s \int_{\vq_1\vq_2\vq_3} g(\vq_1,\vq_2,\vq_3,\vp)\left[\bar\psi_R(\vq_1)\gamma^0\psi_L(\vq_2)\right]\bar\psi_{R}^{\alpha}(\vq_3)
& .
\end{array}
\end{equation}
Due to the Hermiticity of the operator $\cF$, the canonical anticommutation relations are preserved for the transformed $\Psi_{L,R}(\vp)$ fields. With this, as in the bosonic case, it is possible to build a set of scale dependent non-Gaussian fermionic wavefunctionals using \eqref{eq:fermionic_transform} and the Gaussian cMERA for free Dirac fermions given in \cite{cMERA1, cMERA2}.

Being formally guaranteed, it would be interesting to discuss the scale dependent version of these non-Gaussian transformations in a future work.

\section{icMERA: non-Gaussian cMERA Circuits}\label{sec:i_cMERA_circuit}

In this section we first introduce a nonquadratic cMERA entangler to generate non-Gaussian wavefunctionals. Then, we calculate the analytical expressions of the connected part of the two and four point correlators of a generic interacting scalar field theory. To illustrate the proposal, we consider the $\lambda\phi^4$ model and perform the optimization of the tensor network. Finally, we use the optimized variational parameters to evaluate the effective potential and the connected two and four point correlators at various renormalization scales.

\subsection{Nonquadratic Entangler}

Based on the concept of scale dependent nonlinear canonical transformation and non-Gaussian cMERA wavefunctionals exposed above, in this section we propose, as a definition of a nonperturbatively built non-Gaussian cMERA circuit (icMERA), the Hamiltonian evolution produced by the entangler
\begin{align}
\label{eq:icMERA_total_dis}
 K(u)= K_0(u) +  \cB(u)
\ ,
\end{align}
where $K_0(u)$ is the quadratic Gaussian entangler given in  \eqref{eq:entangler} and $\cB(u)$ is 
\begin{align}\label{eq:icMERA_dis}
\cB(u)
=
-s \int_{\vq_1\vq_2\vq_3} g(q_1,q_2,q_3;u) \pi(\vq_1)\phi(\vq_2)\phi(\vq_3)\delta(\vq_1+\vq_2+\vq_3)
\ .
\end{align}
This is a scale dependent operator that nonperturbatively incorporates nonquadratic interaction terms to the cMERA evolution through the variational function $g(q_1,q_2,q_3;u)$, with $s$ being a variational parameter, as in the previous section. 
$g(q_1,q_2,q_3;u)$ is symmetric under exchange of momenta $q_2$, $q_3$ and is parametrized in terms of variational cutoff functions.
 
Thus, as a generalization of the Gaussian cMERA, we define the icMERA evolution operator in the interaction picture as
\begin{align}
\label{eq:icMERA_circuit}
\cU(u_1,u_2)
=
e^{-iu_1 L} \ {\cal P} e^{-i \int_{u_2}^{u_1} ( \hat K_0(u) +  \hat\cB(u) )du}\ e^{i u_2 L}
\ ,
\end{align}
where $\hat K_0(u)$ corresponds to \eqref{disein} and $\hat \cB(u)$ is
\begin{align}\label{eq:icMERA_dis_in}
\hat \cB(u)
=
-s \int_{\vq_1\vq_2\vq_3} g(q_1 e^{-u},q_2e^{-u},q_3e^{-u};u) \pi(\vq_1)\phi(\vq_2)\phi(\vq_3)\delta(\vq_1+\vq_2+\vq_3)
\ .
\end{align}

Having a real wavefunctional requires time reversal invariance of the icMERA evolution which in turn amounts to having an odd number of $\pi$ operators in  $\cB(u)$. In this case $\cB(u)$ is formally equivalent to $\mathcal{B}=\pi\, \phi^2$, \emph{i.e.}, it incorporates cubic interactions into the cMERA evolution. Furthermore, as $s$ in \eqref{eq:icMERA_dis} is a variational parameter related to the coupling strength of the theory,  the standard Gaussian cMERA evolution \eqref{eq:cMERA_evol_free} is recovered when $s \to 0$. 

With this, one may write the icMERA wavefunctional in the Schr\"odinger picture as
\begin{equation}\label{icMERAansatz_ip}
 \Psi[\phi,u]=\langle \phi|\, \cU(u,u_{IR}) | \Omega\rangle
= 
\langle \phi|  \mathcal{P} e^{-i\int^{u}_{u_{\rm IR}} ( K_0(\sigma) +   \cB(\sigma) + L )d\sigma} |\Omega \rangle \, .
\end{equation}

The action of the icMERA operator \eqref{eq:icMERA_circuit} on the field modes $\phi(\vp)$ is given by,
\begin{align}
\cU(0,u)^{-1}\, \phi(\vp)\, \cU(0,u)
=
e^{-iu L}\,  {\cal P} e^{i\int_{u}^{0} (K_0(\sigma) +   \cB(\sigma) )d\sigma}\ \phi(\vp) \ {\cal P} e^{-i\int_{u}^{0} ( K_0(\sigma) +  \cB(\sigma) )d\sigma}\, e^{i u L}
\, .
\end{align}

\subsubsection*{icMERA action on fields} The action of the icMERA operator $K(u) + L$ on $\phi(\vp)$ is given by
\begin{dmath}[spread={3pt}]
	\cU(0,u)^{-1}\, \phi(\vp)\, \cU(0,u)
=
e^{-\frac d2 u }e^{- f(p;u)}\phi(e^{-u}\vp)
+e^{- d u }s\int_{\vq_1\vq_2}f(p,q_1,q_2;u) \frac{e^{- f(p;u)}-e^{-(f(q_1;u)+f(q_2;u))}}{f(p;u)-f(q_1;u)-f(q_2;u)} \phi(e^{-u}\vq_1)\phi(e^{-u}\vq_2)
	\ ,
\end{dmath}
where
\begin{align}
f(p;u)\equiv\int^u_0 d\sigma\ g(pe^{-\sigma};\sigma)
\ ,
\qquad
f(p,q_1,q_2;u)\equiv\int^u_0 d\sigma\, g(pe^{-\sigma},q_1e^{-\sigma},q_2e^{-\sigma};\sigma) \ ,
\label{eq:two-fs}
\end{align}
with $g(p;\sigma)$ given by \eqref{eq:free_var} and having imposed the constraints \eqref{eq:constraint} on $f(p,q_1,q_2;u)$. For convenience, we have dropped the momentum conservation delta $\delta(\vq_1 + \vq_2 -\vp)$ in the integrand. In the same manner, the action on $\pi(\vp)$ can be written as
\begin{dmath}[spread={3pt}]
	\cU(0,u)^{-1}\, \pi(\vp)\, \cU(0,u)
	=
	e^{-\frac d2 u }e^{f(p;u)}\pi(e^{-u}\vp)
	+ 2s e^{-du }\int_{\vq_1\vq_2}f(q_1,q_2,p;u) \frac{e^{ f(p;u)}-e^{f(q_1;u)-f(q_2;u)}}{f(p,u)-(f(q_1;u)-f(q_2;u))} \pi(e^{-u}\vq_1)\phi(e^{-u}\vq_2)
	\ .
\end{dmath}
That is to say, changing the notation to $f(p;u)\equiv f_u(p)$ and $f(p,q_1,q_2;u)\equiv f_u(p,q_1,q_2)$, the scale dependent nonlinear transformations on the fields generated by the icMERA circuit \eqref{eq:icMERA_circuit} are
\begin{align}
\label{eq:nlct_icMERA}
	\begin{aligned}
\cU(0,u)^{-1}\, \phi(\vp)\, \cU(0,u)
	=&\
	e^{-\frac d2 u }e^{-f_u(p)}\phi(e^{-u}\vp)
	\\
	&\
	\hspace{-20pt}
	+e^{- d u }\int_{\vq_1\vq_2}f_u(p,q_1,q_2) \frac{e^{- f_u(p)}-e^{-(f_u(q_1)+f_u(q_2))}}{f_u(p)-f_u(q_1)-f_u(q_2)} \phi(e^{-u}\vq_1)\phi(e^{-u}\vq_2)
	\ ,
	\\[10pt]	
	\cU(0,u)^{-1}\, \pi(\vp)\, \cU(0,u)
	=&\
	e^{-\frac d2 u }e^{ f_u(p)}\pi(e^{-u}\vp)
	\\
	&\
	\hspace{-40pt}
	+2s\ e^{- d u }\int_{\vq_1\vq_2}f_u(q_1,q_2,p) \frac{e^{ f_u(p)}-e^{f_u(q_1)-f_u(q_2)}}{f_u(p)-(f_u(q_1)-f_u(q_2))} \pi(e^{-u}\vq_1)\phi(e^{-u}\vq_2)
	\ .
	\end{aligned}
\end{align}

\subsubsection*{icMERA variational parameters} 

Regarding \eqref{eq:two-fs}, let us note that, in contrast to the function $h$ in \eqref{eq:BBos}, we impose the constraints on the variational functions once the integration over $\sigma$ has been done. Hence, we need to provide a variational ansatz for $g(pe^{-\sigma},q_1e^{-\sigma},q_2e^{-\sigma};\sigma) $ in such a way that $f(p,q_1,q_2;u)$ still satisfies \eqref{eq:constraint}. Based on the example given in \cite{polley89}, we propose
\footnote{We emphasize that other alternative ansatze, based on the mulitple combinations of the cutoff parameters can be also considered.}
\begin{align}
g(p,q_1,q_2;\sigma)
=
g_\cB(\sigma)\Gamma_\cB(p e^\sigma,q_1e^\sigma,q_2e^\sigma)\Gamma\left(\frac{p}{\Lambda }\right)\Gamma\left(\frac{q_1}{\Lambda }\right)\Gamma\left(\frac{q_2}{\Lambda }\right)
\ ,
\label{eq:g-ansatz-phi2}
\end{align}
where $g_\cB(\sigma)$ is the variational parameter that tunes the strength of the scale dependent non-Gaussian transformation, $\Gamma_\cB$ is a combination of cutoff functions depending on the variational cutoff's $\Delta_i$,
\begin{align}
\Gamma_\cB(p,q_1,q_2)
=
\Gamma\left(\frac{p}{\Delta_1}\right)\left[
\Gamma\left(\frac{\Delta_1 }{q_1}\right)
-\Gamma\left(\frac{\Delta_2 }{q_1}\right)
\right]\left[
\Gamma\left(\frac{\Delta_1 }{q_2}\right)
-\Gamma\left(\frac{\Delta_2 }{q_2}\right)
\right]
\ ,
\label{eq:cutoffs}
\end{align}
and $\Gamma\left(q_i/\Lambda \right)$ exhibits the typical cMERA-like variation of the cutoff's with the scale. The optimal function $\Gamma_{\cB}(p,q,r)$ has to be found self-consistently by determining both cutoffs $\Delta_i$. Namely, different from the Gaussian set-up, this scheme illustrates how the strength of the interaction variationally determines the region in momentum space that will be relevant in the optimization procedure. This feature turns out to be essential for strongly-coupled systems, which exhibit some regimes at which the Gaussian quasi-particle picture is no longer valid. In Section \ref{sec:lambdaphi4}, we give evidence of some nonperturbative effects captured by this method.


Integrating over $\sigma$ in \eqref{eq:g-ansatz-phi2}, we obtain
\begin{dmath} 
	f(p,q_1,q_2;u)
	=
	\Gamma_\cB(p,q_1,q_2) \int_0^u
	g_\cB(\sigma)\Gamma\left(\frac{p}{\Lambda e^{\sigma}}\right)\Gamma\left(\frac{q_1}{\Lambda e^{\sigma}}\right)\Gamma\left(\frac{q_2}{\Lambda e^{\sigma}}\right) d\sigma
	\ .
\end{dmath}
This assures that the constraints \eqref{eq:constraint} are satisfied for $f(p,q_1,q_2;u)$ (due to the cutoff combination provided by $\Gamma_\cB$) and we still preserve the structure of a cMERA-like scale transformation where the cutoffs on the momenta are suppressed by a factor $e^{-\sigma}$.

\subsubsection*{Generic non-Gaussian transformation}
When the non-Gaussian entangler $ \cB(u)$ contains $n$ scalar fields (in analogy to $\mathcal{B}=\pi\, \phi^n$ introduced in the previous section),
\begin{eqnarray}
\cB(u)
=
s  \int_{\vk\vq_1\cdots\vq_n}  g(k ,q_1 ,\ldots,q_n ,\sigma) \pi(\vk)\phi(\vq_1)\cdots\phi(\vq_n) \\ \nonumber
\ ,
\label{eq:ng-disentangler-generic}
\end{eqnarray}
(where we have dropped the momentum conservation $\delta(\vk+\vq_1+\cdots+\vq_n)$), then the action of the icMERA operator is given by
\begin{equation}\label{eq:nl_field_trans_icMERA}
	\begin{aligned}
	\cU(0,u)^{-1}\, \phi(\vp)\, \cU(0,u)
	=&\
	e^{-\frac d2 u }e^{-f(p,u)}\phi(e^{-u}\vp)
	\\
	&\
	+e^{- n\frac d2 u }\int_{\vq_1\cdots\vq_n}\tilde h_u(p,q_1,\cdots,q_n) \phi(e^{-u}\vq_1)\cdots \phi(e^{-u}\vq_n)
	\ ,
	\\[10pt]
	\cU(0,u)^{-1}\, \pi(\vp)\, \cU(0,u)
	=&\
	e^{-\frac d2 u }e^{f(p,u)}\pi(e^{-u}\vp)
	\\
	&\
	\hspace{-20pt}
	+n s \ e^{- n\frac d2 u }\int_{\vq_1\cdots\vq_n}\check h_u(q_1,\cdots,q_n,p) \pi(e^{-u}\vq_1)\phi(e^{-u}\vq_2)\cdots \phi(e^{-u}\vq_n)
	\ ,
	\end{aligned}
\end{equation}
where $\tilde h_u$ and $\check h_u$ are defined as
\begin{align}\label{eq:variational_param_icMERA}
\begin{aligned}
\tilde h_u(p,q_1,\cdots,q_n)
\equiv&\
f_u(p,q_1,\cdots,q_n) \ \displaystyle\frac{e^{- f_u(p)}-\prod_{i=1}^n e^{- f_u(q_i)}}{f_u(p)-\sum_{i=1}^n f_u(q_i)}
\ ,
\\ 
\check h_u(q_1,\cdots,q_n,p)
\equiv&\
f_u(q_1,\cdots,q_n,p)\  \displaystyle\frac{e^{ f_u(p)}-e^{ f_u(q_1)}\prod_{i=2}^n e^{ -f_u(q_i)}}{f_u(p)-f_u(q_1)+\sum_{i=2}^n f_u(q_i)}
\ ,
\end{aligned}
\end{align}
and 
\begin{align}
f_u(p,q_1,\ldots,q_n)\equiv\int_0^u  g(pe^{-\sigma},q_1e^{-\sigma},\ldots,q_ne^{-\sigma})d\sigma\, .
\label{eq:f-NG-phin}
\end{align}
which follows the same properties as \eqref{eq:two-fs}. Let us note that the orthogonality constraints of the function $f_u(p,q_1,\cdots,q_n)$ are transferred to the functions $\tilde h_u$ and $\check h_u$. Following the discussions above, for a generic transformation implying \eqref{eq:f-NG-phin}, we straightforwardly generalize our expression for $ g(p,q_1,\cdots,q_n;\sigma)$ as
\begin{align}
g(p,q_1,\cdots,q_n;\sigma)
=
g_\cB(\sigma)\Gamma_\cB(pe^\sigma,q_1e^\sigma,\cdots,q_ne^\sigma) \Gamma\left(\frac{p}{\Lambda }\right) \Gamma\left(\frac{q_1}{\Lambda }\right) \cdots \Gamma\left(\frac{q_n}{\Lambda }\right)
\ ,
\end{align}
where $\Gamma_\cB$ is now given by
\begin{align}
\Gamma_\cB(p,q_1,\cdots,q_n)
=
\Gamma\left(\frac{p}{\Delta_1}\right)\left[
\Gamma\left(\frac{\Delta_1 }{q_1}\right)
-\Gamma\left(\frac{\Delta_2 }{q_1}\right)
\right]\cdots\left[
\Gamma\left(\frac{\Delta_1 }{q_n}\right)
-\Gamma\left(\frac{\Delta_2 }{q_n}\right)
\right]
\ .
\end{align}

The expressions in \eqref{eq:nl_field_trans_icMERA} represent the non-Gaussian version of the scale dependent field transformations \eqref{eq:cMERA_fields} that define the renormalization group flow of a Gaussian cMERA circuit, which can be trivially recovered by simply taking $s \to 0$. As expected, icMERA also works at a perturbative regime. In that respect, a diagrammatic analysis has been carried out in Appendix \ref{app:diagrams} for the $\lambda\phi^4$ theory.

 Finally we would like to make some comments on the structure of the icMERA entangler. The \emph{crosstalking} between the quadratic part $K_0(u)$ and the purely non-Gaussian term $\cB(u)$ is obvious from \eqref{eq:variational_param_icMERA}. In this sense, the quadratic Gaussian part $K_0(u)$ is necessary in order for the wavefunctional in the ${\rm IR}$ to become $ \Psi[\phi,u_{IR}]=\langle \phi|\Omega\rangle$. Thus, while in principle icMERA could be built solely on the pure non-Gaussian part, the Gaussian entangler is fundamental to asymptotically achieve the simple Gaussian state $ |\Omega\rangle$ in the ${\rm IR}$.  

From the point of view of the entanglement flow along an icMERA circuit, as posed in \cite{Cotler1}, the quadratic term is responsible for generating pairwise entanglement between modes as a function of scale \cite{JMV1}, whereas the $\cB(u)$ term generates  $n$-tuplet quantum correlations as a function of scale. How to characterize these higher order quantum correlations is an interesting topic that we leave for future investigations.

\subsection{icMERA Correlation Functions}

Let us focus now on the contribution to the connected correlation functions of a generic interacting scalar field theory when computed by means of the icMERA circuit. This illustrates the \emph{generality} of the trial state posed by an icMERA circuit. That is to say, the icMERA state captures salient physical features such as nonvanishing connected correlation functions in an interacting theory. This gives evidence that our method stands for a solution to the problem of \emph{calculability} mentioned in Section \ref{sec:Intro}. That is to say, an icMERA state constitutes such a flexible \emph{ansatz} for the vacuum wavefunctional that it is easy and straightforward to evaluate expectation values of operators/observables, as for instance correlators. Finally, the multiscale approach given by icMERA provides a procedure to gain some understanding of the nonperturbative effects taking place at different scales. 

From  \eqref{eq:nl_field_trans_icMERA}, one may write the $N$-point correlators at scale $u$ as
\begin{dmath}
G^{(N)}(\vp_1,\cdots,\vp_N;u)
=
\expval{\phi(\vp_1)\cdots\phi(\vp_N)}_{(NG;u)}
\\[10pt]
=
\expval{\tilde U^{-1}(0,u)\, \phi(\vp_1)\, \tilde U(0,u) \cdots \tilde U^{-1}(0,u)\, \phi(\vp_N)\, U^{-1}(0,u)}_{G}\
= \expval{\phi_u(\vp_1)\cdots \phi_u(\vp_N)}_{G}
\\[10pt]
+ s\left [\expval{\bar{\phi}_u(\vp_1)\phi_u(\vp_2)\cdots \phi_u(\vp_N)}_{G} +\cdots +\expval{\phi_u(\vp_1)\cdots \phi_u(\vp_{N-1})\bar{\phi}_u(\vp_N)}_{G}	\right]
\\[10pt]
+ s^2\left[ \expval{\bar{\phi}_u(\vp_1)\bar{\phi}_u(\vp_2)\phi_u(\vp_3)\cdots \phi_u(\vp_N)}_{G}
\\[10pt]
+\cdots +\expval{\phi_u(\vp_1)\cdots\bar{\phi}_u(\vp_{N-1})\bar{\phi}_u(\vp_{N})}_{G}\right]
\\
\mbox{} \, \ \ \vdots
\\
+ s^N \expval{\bar{\phi}_u(\vp_1)\cdots\bar{\phi}_u(\vp_N)}_{G}
\ ,
\end{dmath}
where we have defined
\begin{eqnarray}\label{eq:def_trans_fields_icMERA}
\phi_u(\vk)&\equiv& 	e^{-\frac d2 u }e^{- f(k;u)}\phi(e^{-u}\vk)\, \\ \nonumber
\bar{\phi}_u(\vk)
&\equiv& 	
e^{- n\frac d2 u }\int_{\vq_1\cdots\vq_n}\, \tilde h_u(k,q_1,\cdots,q_n) \phi(e^{-u}\vq_1)\cdots \phi(e^{-u}\vq_n)
\ .
\end{eqnarray}
In other words, the icMERA circuit goes beyond the Gaussian approximation and captures scale dependent nonperturbative contributions for the $N$-th order correlator, which are arranged in powers of the variational parameter $s$. As commented above, in order to quantify to which extent the icMERA ansatz is nonperturbatively characterizing the interactions of the theory under consideration, we have to explicitly calculate the expressions for the connected part of these correlators. In this respect, the connected 2-point and 4-point correlation functions in real space (see also \cite{JJ1}) given by the icMERA circuit with a cubic non-Gaussian entangler $\cB(u)$, \emph{i.e.}, with $n=2$ in \eqref{eq:def_trans_fields_icMERA}, are given by
\begin{equation}
\begin{aligned}
G_c^{(2)}(\va\vb;u)           
&= 
D(\va\vb,u)	+ s^2\, \chi_2 (\va\vb,u) \ ,
\\
G_c^{(3)}(\va\vb\vc;u)           
&= 
s\left[ \chi_3(\va\vb\vc;u)\right]
+s^3\left[ \chi_4(\va\vb,\vb\vc,\vc\va;u)\right]
\ ,
\\
G_c^{(4)}(\va\vb\vc\vd;u) &=\frac{s^2}{2} \left[\chi_5(\va\vb\vc\vd;u)\right] +s^4  \left[\chi_6(\va\vb\vc\vd;u)\right]
\ ,
\end{aligned}
\label{connected_corr}
\end{equation}
where $D(\va\vb;u)$ is 
\begin{align}
D(\va\vb;u)
= 
\frac12 \int_\vp  e^{-2f(p;u)}F(p_u)\, e^{i\vp \cdot \vx_{\va\vb}}
\ ,
\end{align}
and bracketed quantities, which are given in Appendix \ref{app:integrals}, correspond to a series of permutations of the loop integrals\footnote{Here we follow the notation in \cite{polley89, ritschel90}}
\begin{align}
\begin{split}
\chi_2(\va \vb;u) 
=&\
\frac12 \int_{\vp,\vq}  \Upsilon_2(\vp,\vq,u) F(p_u)\, F(q_u)
\ \, e^{i(\vp+\vq)\cdot \vx_{\rm ab}} 
\\ 
\chi_3(\va\vb,\vc\vd;u) 
=&\
\frac12 \int_{\vp\vq} \,  \Upsilon_3(\vp,\vq;u)\,  F(p e^{-u})F(q e^{-u})\, e^{i(\vp\cdot \vx_{\va\vb}+\vq\cdot \vx_{\vc\vd})}
\ ,
\\ \nonumber
\chi_4(\va\vb,\vc\vd,\ve\vf;u) 
=&\
\int_{\vp\vq\vr} \, \Upsilon_4(\vp,\vq;u) F(p e^{-u})F(q e^{-u})F(r e^{-u}) e^{i(\vp\cdot \vx_{\va\vb}+\vq\cdot \vx_{\vc\vd}+\vr\cdot \vx_{\ve\vf})}
\ ,
\\ \nonumber
\chi_5(\va\vb,\vc\vd,\ve\vf;u) 
=&\
\int_{\vp,\vq,\vr}\,  \Upsilon_5 (\vp,\vq,\vr,u)\,  F(p_u)\, F(q_u)\, F(r_u) \ e^{i(\vp\cdot \vx_{\rm ab}+\vq\cdot \vx_{\rm cd}+\vr\cdot \vx_{\rm ef})} \\ 
\chi_6(\va\vb,\vc\vd,\ve\vf,\vg\vh; u) 
=&\
\int_{\vp, \vq, \vr, \vs} \Upsilon_6(\vp,\vq,\vr,\vs,u)\, F(p_u)\, F(q_u)\, F(r_u)\, F(s_u) \\ 
&\ \qquad\qquad \times e^{i(\vp\cdot \vx_{\rm ab}+\vq\cdot \vx_{\rm cd}+\vr\cdot \vx_{\rm ef}+\vs\cdot \vx_{\rm gh})}\, .
\end{split}
\end{align}
We have introduced the variational ``vertices''
\begin{equation}\label{eq:vertex}
\begin{aligned}
 \Upsilon_2(\vp,\vq;u)=&\  c(\vp,\vq;u)^2\, , \\ 
 \Upsilon_3(\vp,\vq;u)=&\  c(\vp,\vq;u)\, , \\ 
  \Upsilon_4(\vp,\vq;u)=&\  c (\vp ,\vq;u)\,  c (\vq ,-\vr;u)\,  c (\vp ,\vr;u)\, , \\ 
 \Upsilon_5 (\vp,\vq,\vr;u)=&\
c(\vp,\vq;u)\,  c(\vp,\vr;u)\, ,\\ 
 \Upsilon_6(\vp,\vq,\vr,\vs;u)=&\
c(\vp,\vq;u)\, c(\vp,\vr;u)\,  c(\vq,\vs;u)\,  c(\vr,\vs;u).
\end{aligned}
\end{equation}
and have used the compact notation ${\va}\equiv\vx_{\rm a}$, ${\va \vb}\equiv\vx_{\rm a b}\equiv\vx_{\rm a} - \vx_{\rm b}$ with $p_u\equiv p\, e^{-u}$. 
The scale dependent variational functions are encoded in $c(\vp,\vq;u)$, which is given by
\begin{align}
c(\vp,\vq;u)
\equiv \tilde h_u (|\vp+\vq|,p,q)\, 
.
\end{align}

It is worth to mention that icMERA, which involves scale dependent nonperturbatively generated wavefunctionals, allows us to study regimes that interpolate between weak and strong coupling. For example, the expressions for the connected parts of the 2- and 4-point functions in \eqref{connected_corr} are fully determined and, depending on the optimal value of $s$, they can receive both perturbative and nonperturbative contributions. As it will be shown in the next section, in the self-interacting scalar $\lambda\, \phi^4$ model, the parameter $s \propto \lambda$. As a result, one may infer from \eqref{connected_corr} that in the perturbative regime, $G_c^{(4)}\sim \cO(\lambda^2)$ while the nonperturbative effects are captured by the term $s^4$.  In this regard, in Appendix \ref{app:diagrams} we provide a perturbative analysis of the connected correlators for the $\lambda\phi^4$ theory.

\subsection{icMERA Circuit for the Scalar $\lambda\, \phi^{4}$ Theory}
\label{sec:lambdaphi4}

In order to fully solve the icMERA tensor network and evaluate the previous expressions for a concrete theory described by a Hamiltonian $\cH$, we must obtain the optimal values for the variational parameters $f(p;u)$, $f(p,q_1,\cdots,q_n;u) $ and $s$. {This is addressed by minimizing the expectation value of the energy density w.r.t the icMERA ansatz for a fixed length scale $u$, \emph{i.e.}, $\langle \cH \rangle_u = \langle \Psi_u |\, \cH\,  | \Psi_u\rangle\,$}. {Our aim here is to obtain the optimized parameters for an icMERA tensor network circuit representing the ground state of the self-interacting $\lambda  \phi^4$ scalar theory in (1+1) dimensions. The Hamiltonian density for this model reads 
\begin{align}
\cH =\f12\left(\pi(x)^2 + \left(\nabla \phi(x)\right)^2\right) + \frac{1}{2}\ m^2\, \phi(x)^2 + \frac{\lambda}{4!}\phi(x)^4\, ,
\end{align}	
where $m$ and $\lambda$ are the bare mass and the bare coupling of the theory respectively. The $\lambda \phi^4$  scalar field theory in two dimensions provides an example of a nontrivial interacting field theory. According to \cite{chang1}, this model experiences a second order phase transition at which the vacuum changes continuously from a symmetric to a nonsymmetric state. However, the rigorous proof \cite{mcbryan} of this fact does not allow to compute the critical coupling. An estimate was obtained by the variational Gaussian approximation \cite{chang2}, but this yields a wrong critical behavior as it predicts a first order phase transition.

We will consider the icMERA given by a $\pi\, \phi^2$ kind of nonquadratic interaction term in the entangler $K(u)$. With the icMERA ansatz given in \eqref{icMERAansatz_ip},  and taking into account the following correlators when evaluated at the same point $x$,
\begin{equation}\label{eq:phi_ng}
\begin{aligned}
\expval{\phi(x)}_u =&\ \chi_0 + s \chi_1(u)\equiv \phi_c  \, ,\\ 
\expval{\phi^2(x)}_u =&\ I(u) + \phi_c^2 + s^2\, \chi_2(u)\, , \\ 
\expval{\phi^4(x)}_u =&\ 3I(u)^2+6s^2(I(u)\chi_2(u)+\chi_5(u)) +3s^4(\chi_2(u)^2+ \chi_6(u))
\\
&\ +4\phi_c(3s\, \chi_3(u)+s^3\, \chi_4(u))+6\phi_c^2(I(u)+s^2 \chi_2(u))+\phi_c^4 \, ,
\end{aligned}
\end{equation}
the expectation value of the energy functional for the $\lambda \phi^4$ theory is given by
\begin{dmath}
\expval{\cH}_u
=\expval{\cH_{\rm kin}}_u +\frac12 m^2(s^2 \chi_2(u)+\phi_c^2) 
+\frac{\lambda}{4!}\left[ 3I(u)^2+6s^2(I(u)\chi_2(u)+\chi_5(u)) +3s^4(\chi_2(u)^2+ \chi_6(u))
+4\phi_c(3s\, \chi_3(u)+s^3\, \chi_4(u))+6\phi_c^2(I(u)+s^2 \chi_2(u))+\phi_c^4 
\right]  \, ,
\label{sm-eq:energy_functional_phi4}
\end{dmath}
with 
\begin{align}
\expval{\cH_{\rm kin}}_u=\frac{1}{4}\int_{\vp}\left[F(p;u)^{-1} + p^2F(p;u)\right] + s^2 \chi_7(u)\, ,
\end{align}
where $F(p;u)$ is given in \eqref{dictionaryJMV}. Here, the momentum integrals $\chi_i(u)$, which are given in Appendix \ref{app:integrals}, here are evaluated at the same spatial point $x$ and 
\begin{align}
I(u) \equiv \f12 \int_\vp F(p;u)\, .
\end{align}

The variational terms in the integrals $\chi_i$ can be understood as a kind of generalized condensates yielded by the icMERA ansatz. To see this, we note that, for small $s$, the term  $\sim \phi_c\, \chi_3(u)$ in \eqref{sm-eq:energy_functional_phi4} is the major contribution to the improvement of the energy value compared to the Gaussian estimate (which is obtained when taking the limit $s\to 0$ in \eqref{sm-eq:energy_functional_phi4}) \cite{polley89, ritschel90, JJ1}. This term is formally equivalent to the one coming from the interaction with a background field. In a Gaussian ansatz, that background field is given by $\chi_0$ in \eqref{cMERAwavefunctional1shifted}. Indeed, it has been shown that the optimal $\chi_3(u)$ contains an infinite series of contributions to the two-point function that correspond to the ``cactus''-diagrams resummation obtained from a pure Gaussian ansatz\cite{ritschel94}. This is in agreement with understanding $\chi$'s in \eqref{sm-eq:energy_functional_phi4} as a series of generalized non-Gaussian condensates. 

In this sense, in some limits the Gaussian wavefunctional \eqref{cMERAwavefunctional1shifted} turns out to be a special case of the $\pi \phi^2$-kind of wavefunctionals yielded by icMERA \cite{ritschel90}. The idea is to define $\mathbb{P}$ and $\mathbb{Q}$ as the domain supports for the transformed $p$-modes of the field and the support for the shifting $q$-modes respectively (see \eqref{eq:nlct_icMERA}).  $\mathbb{P}$ is a sphere with volume ${\rm V}_{\mathbb{P}}$ and center at the origin and $\mathbb{Q}$ is a spherical shell which surrounds $\mathbb{P}$ with volume ${\rm V}_{\mathbb{Q}}$. Taking the limit of small $\varepsilon \equiv  {\rm V}_{\mathbb{P}}/{\rm V}_{\mathbb{Q}}$ it is possible to show that the only ``condensate'' independent of $\varepsilon$ is $\chi_1$ while for the remaining $\chi$'s one can easily obtain upper bounds which depend on $\varepsilon \to 0$ \cite{ritschel90}. By considering $\phi_c$ as a new parameter, the final energy expectation value may be obtained directly from the Gaussian result by substituting $\chi_0 \to \phi_c$. Again, this strongly suggests that the parameters $\chi$ act as a kind of ``higher order'' non-Gaussian condensates that expand the ability of the ansatz to improve the variational estimation of ground state energy. 

\subsubsection*{Optimization} 
The optimal values of the variational parameters $f(p;u)$ and $f(p,q,r;u)$ can to be found by setting 
\begin{equation}
\label{eq:optimization}
\frac{\delta\, \expval{\cH}_u}{\delta f(p;u)}=0\, , \quad  \frac{\delta\, \expval{\cH}_u}{\delta c(\vp,\vq;u)}=0\, .
\end{equation}
 Despite this can be done in full generality, $\phi_c$ has to be fixed, in order for the trial wavefunctions to be consistent with the Rayleigh-Ritz method \cite{sm-Stevenson:1985zy,sm-Sher:1988mj}. In other words, one has to fix $\phi_c$, and minimize with respect to the rest of the variational parameters. In this form, this yields a set of nonlinear coupled equations that must be solved numerically and self-consistently. These can be found in the appendices of \cite{JJ1}. As noted there, the product $s\,  c(\vp,\vq;u)$ is really meaningful and one is allowed to fix
 \begin{align}
s = -4\lambda\, \phi_c\, ,
 \end{align}   
 as a way to conveniently normalize $c(\vp,\vq;u)$. With this at hand, here we will focus on the solution of these equations when $\phi_c=0$, a value for which the optimization equations  greatly simplify. On one hand, the optimization parameter $f(p;u)$ reduces to the Gaussian solution
\begin{align}
f(p,u)
=
\left\{
\begin{array}{llll}
\displaystyle
\frac{1}{4}\log \frac{\Lambda^2e^{2u} + \mu^2}{\omega_{\Lambda}^2}&, \ \ \ & p\leq\Lambda e^{u} &,
\\
\displaystyle
\frac{1}{4}\log \frac{p^2+\mu^2}{\omega_{\Lambda}^2} &, \ \ \  & p>\Lambda e^{u}& ,
\end{array}
\right.
\label{eq:f-optimized}
\end{align}

where $\mu^2$ is a variational mass term given by
\begin{equation}
\label{gapeqn_icMERA}
\mu^2 = m^2 +\frac{\lambda}{2}\, I(\mu^2)\,.
\end{equation}
Thus, the variational parameter appearing in the quadratic entangler $K_0(u)$ of icMERA is 
\begin{align}
\label{g_quadratic}
g(u) =& \, \frac{1}{2} \, \frac{\Lambda^2\, e^{2u}}{\mu^2 + \Lambda^2\, e^{2u}} \, .
\end{align}
 
Because $c(\vp,\vq;u_{UV})=c(\vp,\vq;u_{IR})=0$, we need to solve the optimization equation for $c(\vp,\vq;u)$ at a particular scale. Then it is rather convenient to take the scale $u_{*} = \log \left(\mu/\Lambda\right)$ in order to solve the equation for the variational parameter $c(\vp,\vq;u)$. For an arbitrary scale and to order $\cO(\mu/\Lambda)$, we obtain
\begin{align}
f(\vp,\vq;u)
=
\frac{\Gamma_\cB(|\vp+\vq|,p,q)}{F^{-1}(|\vp+\vq|;u)(F^{-1}(p;u)+F^{-1}(q;u))+|\vp+\vq|^2+\mu^2}
\ .
\end{align}

On the other hand, recalling \eqref{eq:two-fs}, and defining
\begin{align}
f(\vp,\vq;u)\equiv f(|\vp+\vq|,p,q;u)
\ ,
\end{align}
we may write
\begin{dmath}
	f(\vp,\vq;u)
	=
	\int_0^u g(|\vp+\vq| e^{-\sigma},p e^{-\sigma},q e^{-\sigma})d\sigma
	=
	\int_0^u g_\cB(\sigma)\Gamma_\cB(|\vp+\vq|,p,q)\Gamma\left(\frac{|\vp+\vq|}{\Lambda}\right) d\sigma
	=
	\Gamma_\cB(|\vp+\vq|,p,q) \int_0^{-\log(\Lambda/|\vp+\vq|)} g_\cB(\sigma) d\sigma
	\ .
\end{dmath}
Then, for $|\vp+\vq|\to \Lambda e^u$, and for the case in which $f(p;u)\ll1$, we have
\begin{align}
\left.\frac{\partial f(\vp,\vq;u)}{\partial p}\right|_{|\vp+\vq|=\Lambda e^{u}}
=
\left.\frac{\partial f(\vp,\vq;u)}{\partial q}\right|_{|\vp+\vq|=\Lambda e^{u}}
=
\frac{e^{-u} g_\cB(u)}{ \Lambda}
=-\frac{ \Lambda  e^u}{\left(3 \mu ^2+ \Lambda ^2 e^{2 u}\right)^2}\ ,
\end{align}
so the scale dependent variational parameter asssociated to the non-Gaussian entangler $\cB(u)$ in icMERA is given by 
\begin{align}
g_\cB(u)= - \displaystyle\frac{ \Lambda ^2 e^{2 u} }{\left(\frac34 \mu ^2+ \Lambda ^2 e^{2 u}\right)^2}
\ .
\end{align}
This implies that
\begin{align}
f(\vp,\vq;u)
=
\left\{
\begin{array}{llll}
\displaystyle\frac{\Gamma_\cB(|\vp+\vq|,p,q)}{F^{-1}(\Lambda e^u)\left[
	F^{-1}(\Lambda e^u)
	+F^{-1}(\Lambda e^u)
	\right]
	+\Lambda^2 e^{2u}+\mu^2}
&,&|\vp+\vq|\leq \Lambda e^u&,
\\[15pt]
\displaystyle\frac{\Gamma_\cB(|\vp+\vq|,p,q)}{F^{-1}(|\vp+\vq|)(F^{-1}(p)+F^{-1}(q))+|\vp+\vq|^2+\mu^2}
&,&|\vp+\vq|> \Lambda e^u&.
\end{array}
\right.
\end{align} 

To fully solve the optimization process, we need to find the optimal values of the variational cutoffs $\Delta_1$ and $\Delta_2$ inside $\Gamma_\cB$ for a given coupling. This can be achieved by plugging the optimal expressions for $f(p;u)$ and $f(\vp,\vq;u)$ into \eqref{sm-eq:energy_functional_phi4} which leaves $\expval{\cH}_u$ only as a function of  $\Delta_1$ and $\Delta_2$. With this, it is straightforward to perform a direct numerical search for values that minimize \eqref{sm-eq:energy_functional_phi4}.

\subsubsection*{Correlation Functions}
As mentioned above, once the optimal variational parameters of the icMERA-$\pi\phi^2$ circuit, $f(p,u)$ and $f(\vp,\vq;u)$, are obtained for the $\lambda\, \phi^4$ theory, then higher-order correlation functions can be computed through equations \eqref{connected_corr}. As commented previously, the knowledge of higher order correlation functions is necessary for distinguishing the ground states of an interacting system from of a noninteracting one.  With the aim to illustrate the performance of an optimized icMERA circuit, we have carried out computations of the connected part of the two and four point correlation functions in \eqref{connected_corr} for the $\lambda\, \phi^4$ theory. To this end we have optimized the icMERA circuit for this model under the prescriptions given above for different values the interaction strength. For the range of interaction strenghts that have been considered, the results were $\Delta_1 \approx (3/50)\, \Delta_2$ and $\Delta_2 \approx  \Lambda$. These values present exhibit some slight changes for the range of $\lambda$ that we considered.

\begin{figure}[t]
	\centering
	\begin{tabular}{cccc}
		\includegraphics*[width=0.25\textwidth]{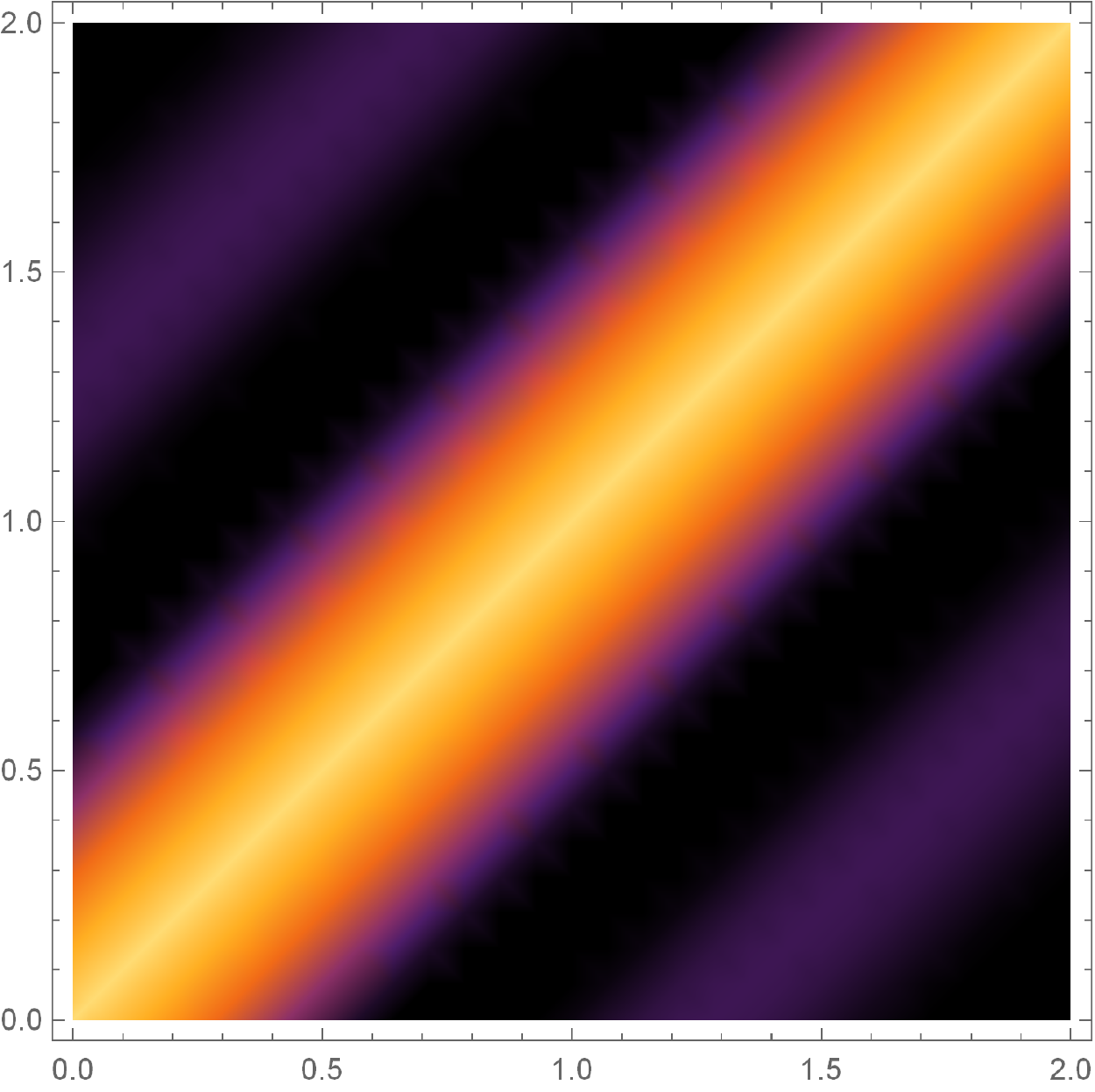}&
		\includegraphics*[width=0.25\textwidth]{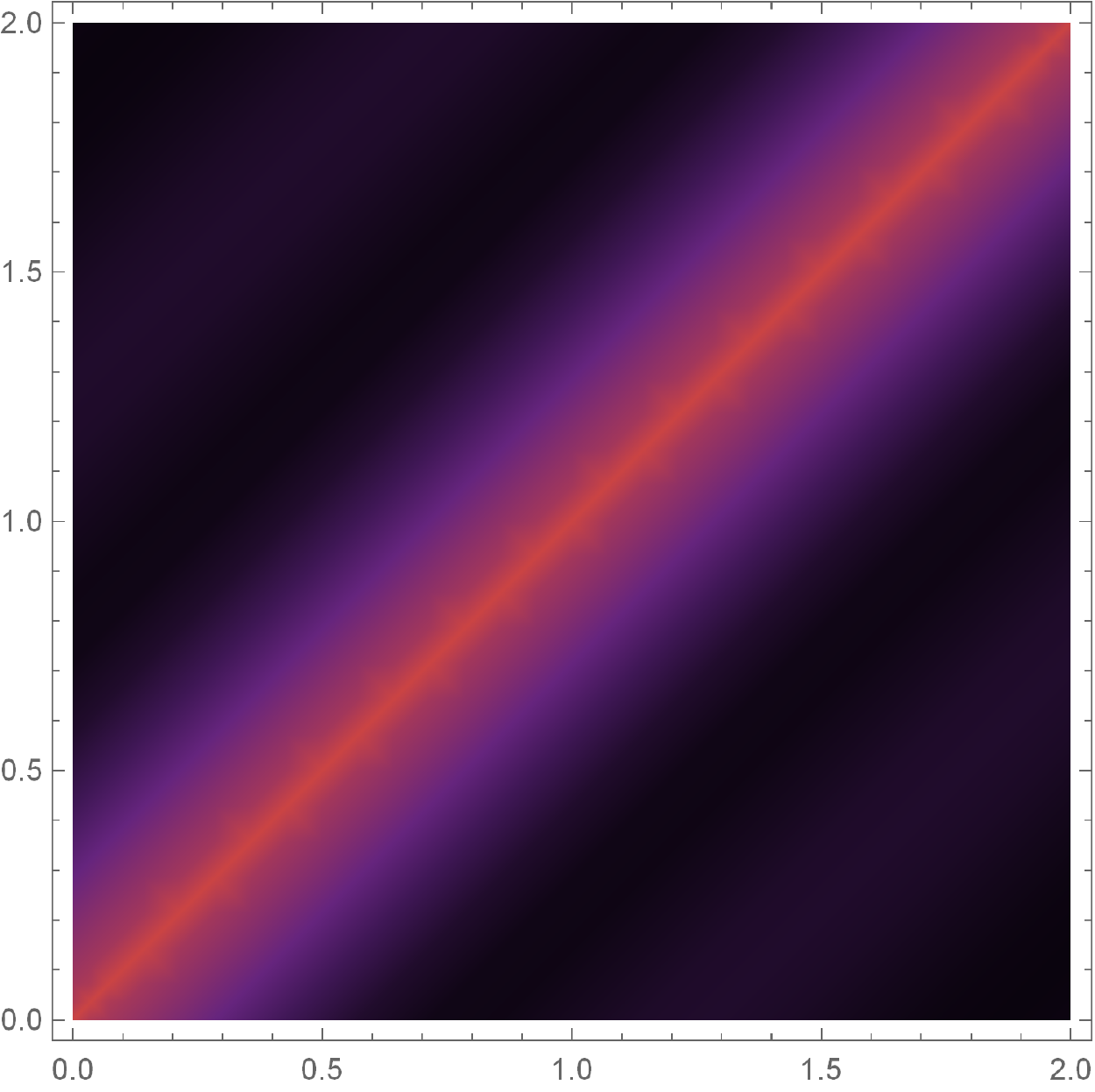}&
		\includegraphics*[width=0.25\textwidth]{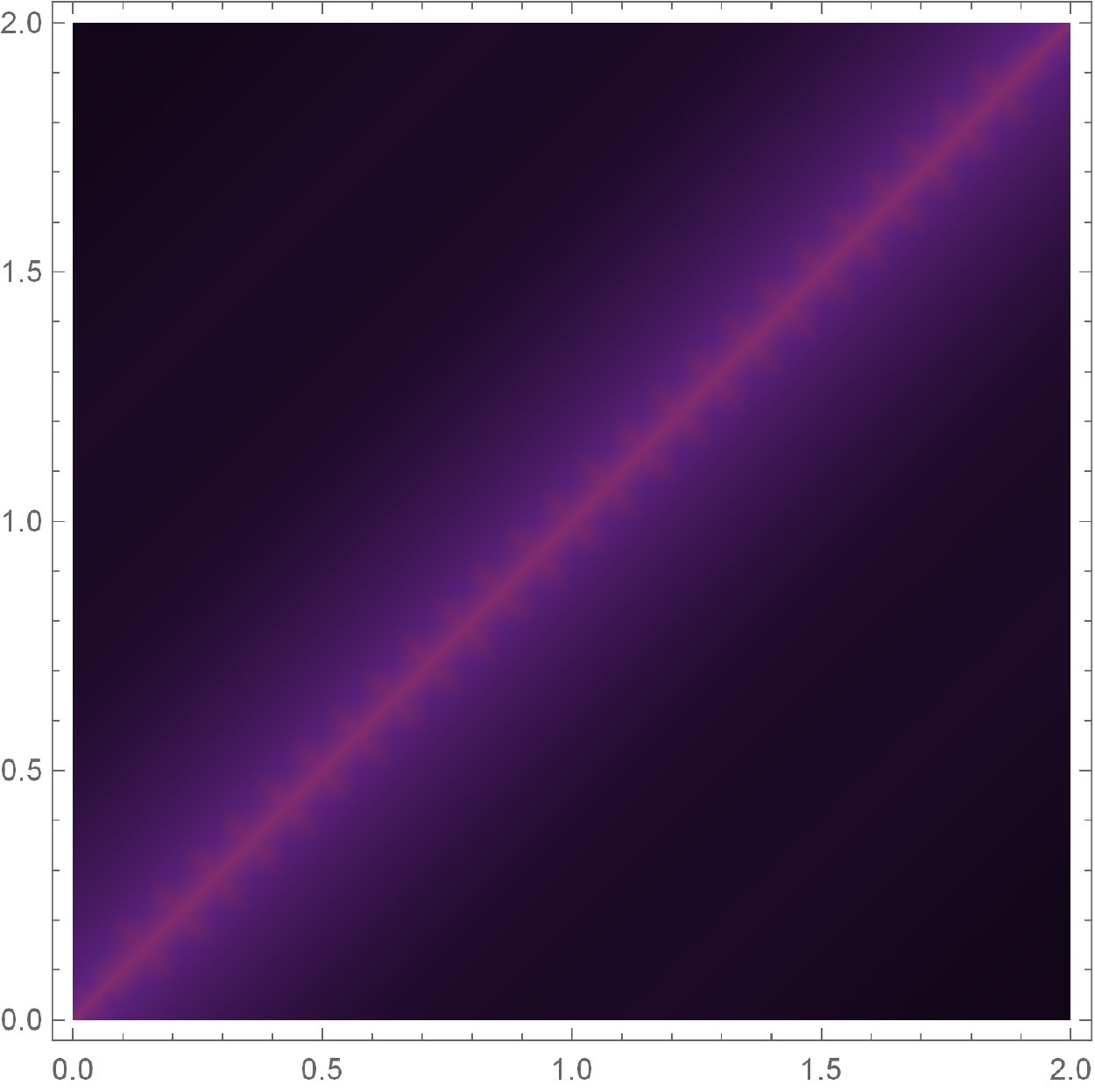}&
		\includegraphics*[width=0.045\textwidth]{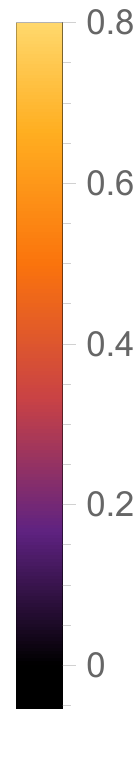}\\
		\includegraphics*[width=0.25\textwidth]{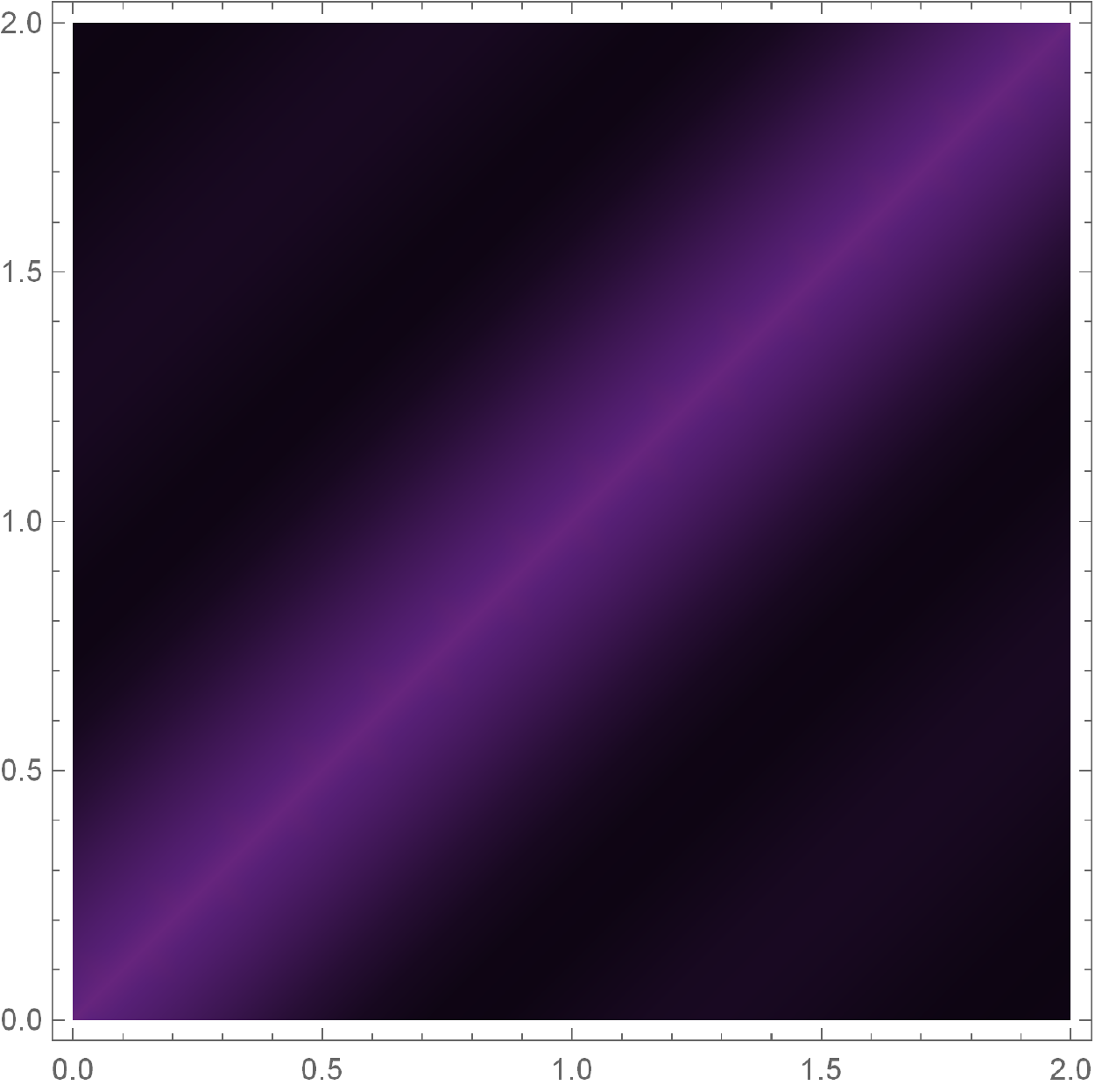}&
		\includegraphics*[width=0.25\textwidth]{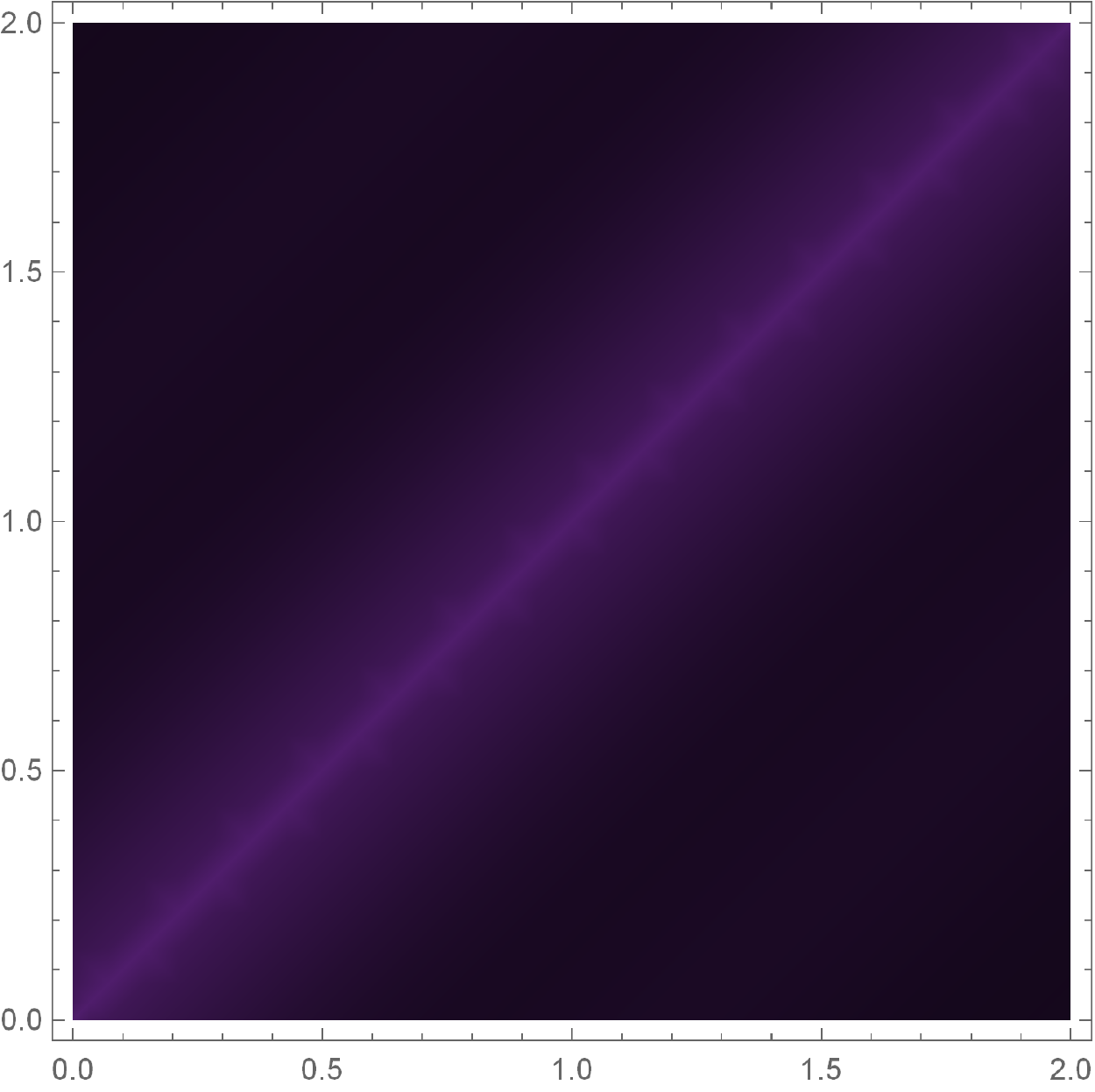}&
		\includegraphics*[width=0.25\textwidth]{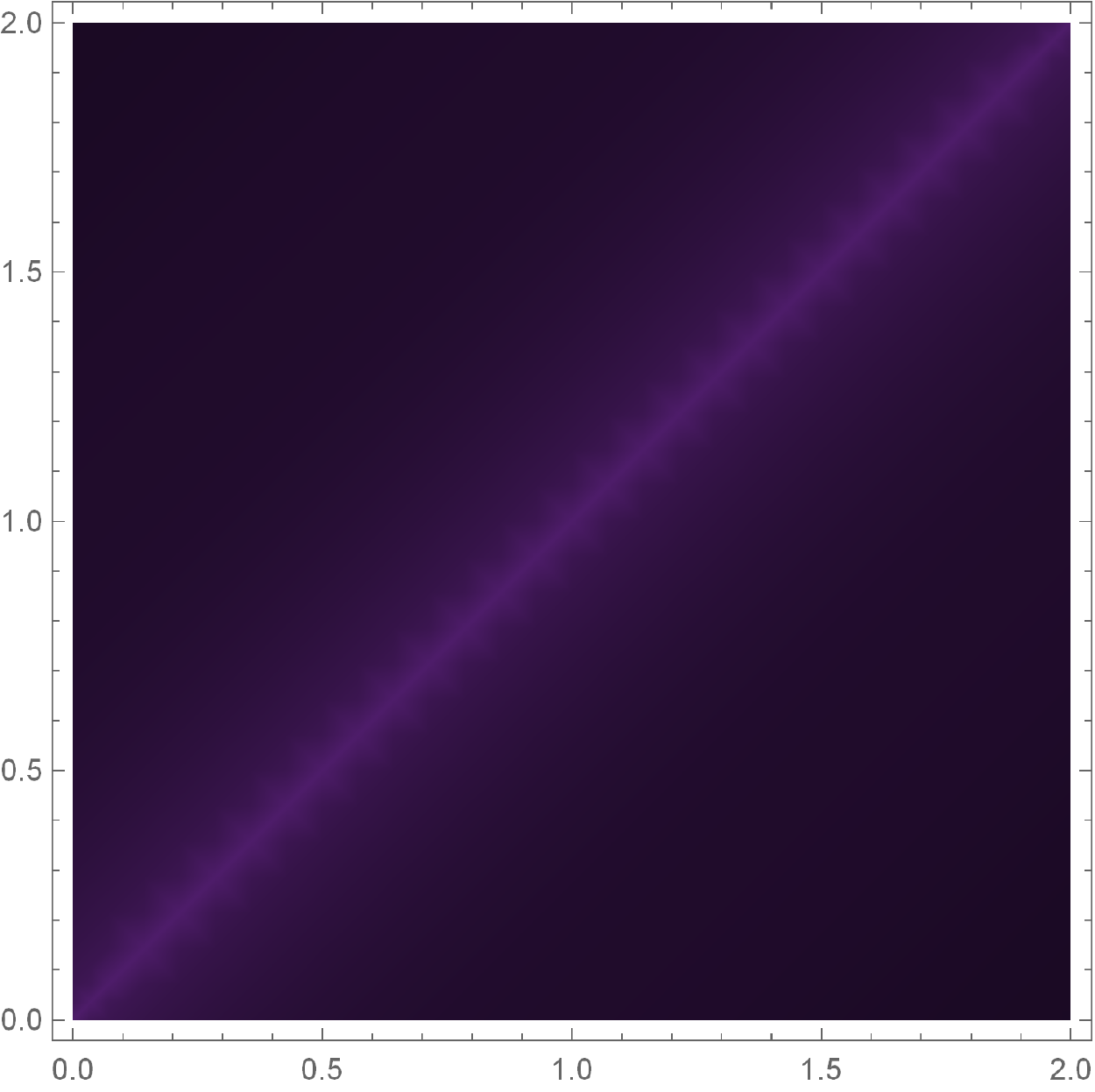}&
		\includegraphics*[width=0.055\textwidth]{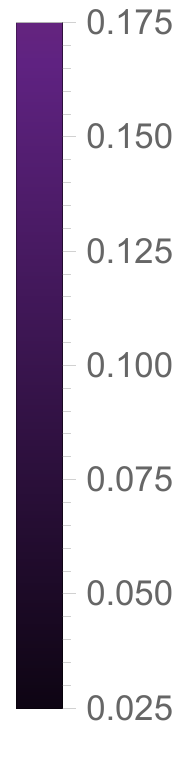}		
	\end{tabular}
	\caption{\textit{\small Density plots of 2-point connected correlation function 
			for the ground state of the $\phi^4$ theory. The interaction 
			strength varies from $\lambda=3.5$ (\emph{left}), $\lambda=2.5$ (\emph{center}) 
			and $\lambda=1.5$ (\emph{right}). \emph{Top} $\sigma = 0.1\, \Lambda$. \emph{Bottom}, $\sigma = 0.01\, \Lambda$. 
			Parameters in the calculation are given by $m=1$, $\phi_c = 1/\sqrt{8}$ and $\Lambda = 100$.}}
	\label{figure:2pt_corr_1}
\end{figure}

Figure \ref{figure:2pt_corr_1} shows the connected part of the two point correlation function $G_c^{(2)}(\vx_1,\vx_2;u)$ in \eqref{connected_corr} at different length scales. Taking $u=\log \sigma /\Lambda$ with $\sigma$ an arbitrary mass scale and an invariant probing distance regime given by $L = 2$, in the first row of the figure it is shown $G_c^{(2)}$ at a length scale labeled by $\sigma = 0.1 \Lambda$, \emph{i.e.}, the 2-point correlation between  coarse grained sites  $\tilde \vx_1 = \vx_1 e^{u}$ (horizontal axis) and $\tilde \vx_2e^{u}$ (vertical axis) ranging from $[0,L]$ in units of $ (0.1 \Lambda)^{-1}$. Thus, we are probing the correlations at an intermediate scale still far from the IR where the non-Gaussian features of the interactions are noteworthy. The second row  shows the 2-point correlations between coarse grained sites ranging from $[0,L]$ in units of the lattice spacing $(0.01\,\Lambda)^{-1}$, \emph{i.e.}, the correlations at very large distances. The different columns show that an increasing of the interaction coupling results in a striking growth of the ground state correlations that are also longer in range than in the free case. 

\begin{figure}[t]
	\centering
	\begin{tabular}{cccc}
		\includegraphics*[width=0.085\textwidth]{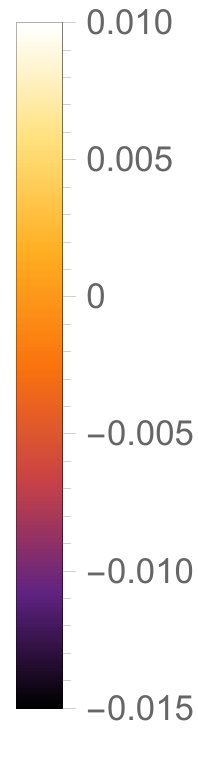}&
		\includegraphics*[width=0.35\textwidth]{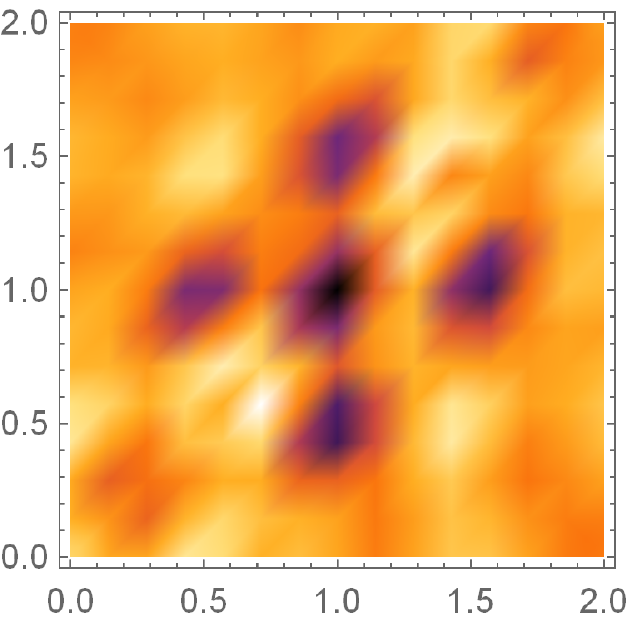}&
		\includegraphics*[width=0.35\textwidth]{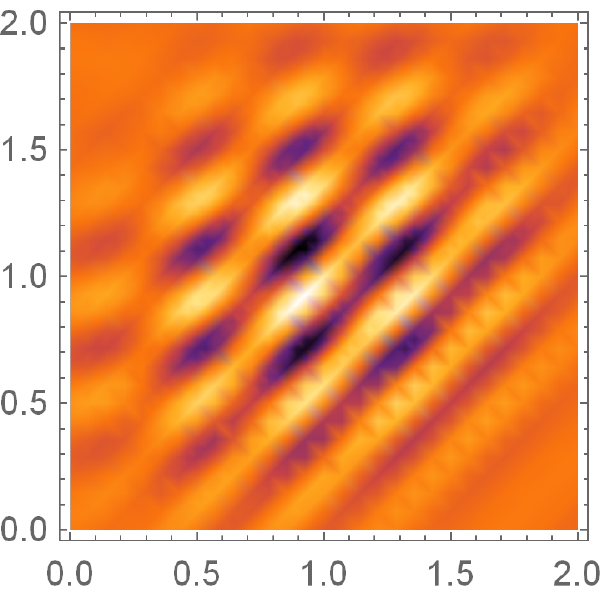}&
		\includegraphics*[width=0.075\textwidth]{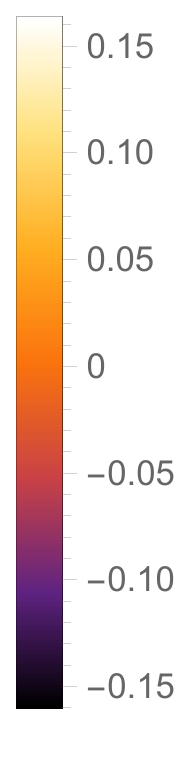}\\
	\end{tabular}
	\caption{\textit{\small Density plot of the two contributions to the 4-point connected correlation function for $\lambda =2$ and $\sigma=0.3.$}}
		\label{figure:4pt_contributions}
\end{figure}

Figure \ref{figure:4pt_contributions} shows the real space structure of the two contributions to the 4-point connected correlation function $G_c^{(4)}(\va\vb\vc\vd;u)$. The left plot of the figure represents the term $\frac{s^2}{2}\, \left[\chi_5\right]$, which is dominant for small coupling strength $\lambda$. In the right part it is represented $s^4\,  \left[\chi_6\right]$, which is the dominant part for larger coupling strengths. In order to visualize the high dimensional data, we choose $\va$ (horizontal axis) and $\vb$ (vertical axis) as points ranging from $[0,L]$ in units of $\Lambda^{-1}$ (\emph{i.e.}, we show the structure of these contributions in the deep UV regime), while  $\vc=L/4$ and  $\vd=3L/4$ are fixed. Both contributions show nonvanishing correlations along the antidiagonal section $\vb = L-\va$ which signals the non-Gaussianity of the state. 

\begin{figure}[t]
	\centering
	\begin{tabular}{ccc}
		\includegraphics*[width=0.30\textwidth]{chi5LU.pdf}&
		\includegraphics*[width=0.30\textwidth]{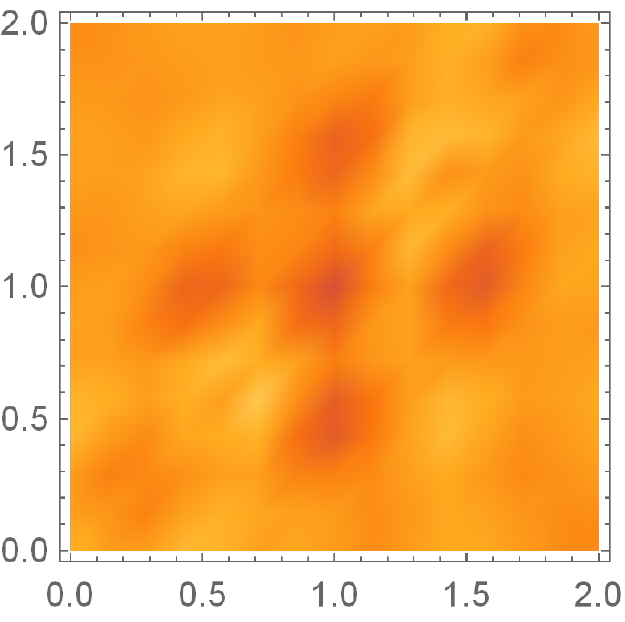}&
		\includegraphics*[width=0.075\textwidth]{chi5_scale}\\
		\includegraphics*[width=0.30\textwidth]{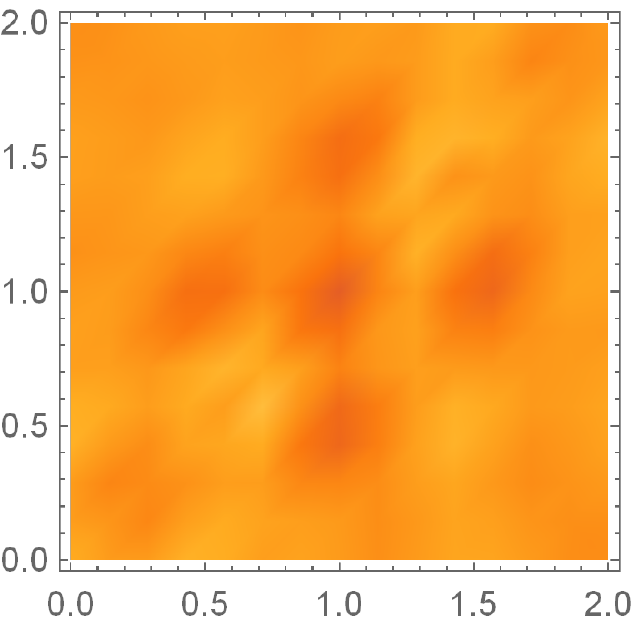}&
		\includegraphics*[width=0.30\textwidth]{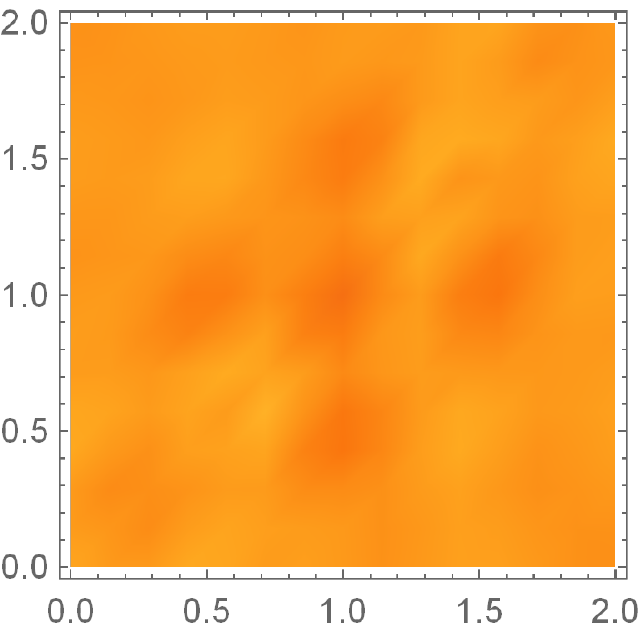}&
	\end{tabular}
	 \caption{\textit{\small Density plots of $s^4 \left[ \chi_5\right]$. 
         (\emph{Top}) $\sigma = 0.3\, \Lambda$\, (\emph{Bottom}) $\sigma = 0.07\, \Lambda$.  \emph{Left} $\lambda=2.0$, \emph{Right}  $\lambda=1.0$.}}
	\label{figure:4pt_corr_1}
\end{figure}

Finally, in Figure \ref{figure:4pt_corr_1} we show the real space structure of the $s^2$-contribution to the 4-point connected correlations for different interaction strengths and length scales. The value of these correlations dramatically diminishes for mass scales smaller than $\sigma \sim 0.01 \Lambda$, which is in agreement with the expected cMERA \emph{Gaussianization} of the state as one approaches the deep IR regime. Our results also show that icMERA is able to capture proper and characteristic scale dependent information about the system at larger orders than those provided by the Gaussian ansatz. Such scale dependence occurs because of the variationally optimized ``vertices'' $\Upsilon_5 $ and $\Upsilon_6$ in Eq. \eqref{eq:vertex}.

\section{Discussion}\label{sec:discussion}
The non-Gaussian icMERA circuit presented in this work introduces a new variational tool to address strongly interacting field theories by means of a systematic building of non-Gaussian wavefunctionals. In this sense, icMERA provides a tool to study how the structure of interactions are  encoded in the correlations and the entanglement patterns of the wavefunctionals of interacting field theories. In our proposal, we have shown that the icMERA circuit allows us to probe, at different length scales, the genuinely non-Gaussian structure of the interactions in a theory.} Regarding the aforementioned entanglement patterns, those are well understood for the case of free theories in terms of the RG flow implemented by the Gaussian cMERA. In the case of interacting theories, it is expected that non-Gaussian correlations establish more complex patterns of entanglement at different length scales. Thus, it would be very interesting to carry out a systematic study of these quantum correlations in future works.  

Taking into account the predictability of our method (which is obviously based on its calculability), it would be interesting to elucidate to what extent the MERA discrete tensor network is able to capture genuinely non-Gaussian features (as for example, connected correlators) when applied to interacting theories.

On the other hand, the multiscale approach provided by an icMERA circuit, may be useful to address recent experimental data on higher order correlation functions in many body systems \cite{kurtosis, schied_nature}. As a fact, despite the most fundamental laws of physics are usually explored in experiments probing the smallest distances, it has been recently shown that models designed to give account of the observations in these high energy experiments, may emerge as effective descriptions of many-body systems at lower energies, \emph{e.g.} in condensed matter physics and quantum simulation experiments. Hence, it is worth to investigate how an icMERA circuit could give account of these data by fixing the laboratory cutoff and the scale energy at which the experiment is performed.

This actually connects to the fact that it is not clear that our ansatz captures the full nonperturbative structure of the ground state of the $\phi^4$ model, which is a very interesting question to be addressed in a near future.

Finally, having such a robust prescription to address the entanglement renormalization of interacting theories at a nonperturbative level, we expect to unveil in a near future the concrete holographic realizations that icMERA is able to exhibit. For example, it would be interesting to explore the complexity of QFT from a cMERA point of view \cite{Chapman:2017rqy,HolCMERA4,Jefferson:2017sdb,Bhattacharyya:2018bbv}.

\section*{Acknowledgments}
JMV acknowledges J. Cotler for useful comments on this manuscript.
The work of JJFM is supported by Universidad de Murcia-Plan Propio Postdoctoral, the Spanish Ministerio de Econom\'ia y Competitividad and CARM Fundaci\'on S\'eneca under grants FIS2015-28521 and 21257/PI/19.
JMV is funded by Ministerio de Ciencia, Innovaci\'on y Universidades PGC2018-097328-B-100 and Programa de Excelencia de la Fundaci\'on S\'eneca Regi\'on de Murcia 19882/GERM/15.

\appendix

\section{icMERA at a Perturbative Regime in $\lambda\phi^4$}
\label{app:diagrams}

In this appendix, we carry out an analysis of the connected part of the correlation functions yielded by an icMERA circuit in the weak interaction regime. We compare the well known perturbative diagrammatic expansions of the $\lambda \phi^4$ theory for this correlators with the diagrammatic structure  associated to an icMERA circuit.

\subsection{Connected diagrams for $G^{(2)}_c$ and $G^{(4)}_c$} 
The lowest contributions to the perturbative diagramatic expansion of the two and four-point functions of the $\lambda \phi^4$ theory in the normal phase of the system (e.g, $\phi_c = 0$) are given in terms of the following Feynman diagrams (see, for example \cite{Ramond:1981pw}):
\begin{dmath}
\label{eq:2p-feynman}
\resizebox{3cm}{!}{\begin{tikzpicture}[baseline={(current bounding box.center)},scale=7]
\begin{feynman}
\diagram* {
a --  b  --  c [blob] --  d --  e,
};
\end{feynman}
\end{tikzpicture}}
\quad
=
\quad
\resizebox{3cm}{!}{\begin{tikzpicture}
\begin{feynman}
\diagram* {
a --  b  --  c  --  d --  e,
};
\end{feynman}
\end{tikzpicture}}
\quad
+
\quad
\resizebox{3cm}{!}{\begin{tikzpicture}
\begin{feynman}
\vertex (a);
\vertex [right=of a] (c);
\vertex [right=of c] (d);
\vertex [above=of c] (c1);
\vertex [above=of c1] (c2);
\diagram* {
(a) -- (c) -- (d) ,
};
\path (c)--++(90:0.3) coordinate (A);
\draw (A) circle(0.3);
\end{feynman}
\end{tikzpicture}}
\quad
+
\quad
\resizebox{3cm}{!}{\begin{tikzpicture}[baseline={(current bounding box.center)}]
\begin{feynman}
\vertex (a);
\vertex [right=of a] (c);
\vertex [right=of c] (d);
\vertex [above=of c] (c1);
\diagram* {
(a) -- (c) -- (d) ,
};
\draw (c) circle(0.3);
\end{feynman}
\end{tikzpicture}}
\quad
+
\quad
\resizebox{3cm}{!}{\begin{tikzpicture}
\begin{feynman}
\vertex (a);
\vertex [right=of a] (c);
\vertex [right=of c] (d);
\vertex [above=.9 of c] (c1);
\diagram* {
(a) -- (c) -- (d) ,
};
\path (c)--++(90:0.3) coordinate (A);
\draw (A) circle(0.3);
\path (c1)--++(90:0) coordinate (A1);
\draw (A1) circle(0.3);
\end{feynman}
\end{tikzpicture}}
\quad
+
\quad
\resizebox{3cm}{!}{\begin{tikzpicture}
\begin{feynman}
\vertex (a);
\vertex [right=of a] (b);
\vertex [right=of b] (c);
\vertex [right=of c] (d);
\vertex [right=of d] (e);
\diagram* {
(a) -- (b) -- (c) -- (d) -- (e),
};
\path (b)--++(90:0.6) coordinate (b1);
\draw (b1) circle(0.6);
\path (d)--++(90:0.6) coordinate (d1);
\draw (d1) circle(0.6);
\end{feynman}
\end{tikzpicture}}
\quad
+
\quad
\cO(\lambda^3)
\ .
\end{dmath}

The diagrams for the connected part of the 4-point function are depicted as 
\begin{dmath}
\label{eq:4p-feynman}
\resizebox{2cm}{!}{\begin{tikzpicture}[baseline={(current bounding box.center)}]
\begin{feynman}[small]
\vertex [blob](v) {};
\vertex [above right=of v] (f1);
\vertex [ below right=of v] (f2);
\vertex [above left=of v] (i1);
\vertex [below left=of v] (i2);
\diagram* {
(i1) --  (v) [blob] --  (i2),
(f2) --  (v),
(f1) --  (v),
};
\end{feynman}
\end{tikzpicture}}
\quad
=
\quad
\resizebox{2cm}{!}{\begin{tikzpicture}[baseline={(current bounding box.center)}]
\begin{feynman}[small]
\vertex (v) ;
\vertex [above right=of v] (f1);
\vertex [ below right=of v] (f2);
\vertex [above left=of v] (i1);
\vertex [below left=of v] (i2);
\path (v)--(f1) coordinate[pos=0.5] (e);
\coordinate (h) at ($(e)+({1/sqrt(2)},-{1/sqrt(2)})$);
\diagram* {
(i1) --  (v) [blob] --  (i2),
(f2) --  (v),
(f1) --  (v),
};
\end{feynman}
\end{tikzpicture}}
\quad
+
\quad
\resizebox{2cm}{!}{\begin{tikzpicture}[baseline={(current bounding box.center)}]
\begin{feynman}[small]
\vertex (v) ;
\vertex [above right=of v] (f1);
\vertex [ below right=of v] (f2);
\vertex [above left=of v] (i1);
\vertex [below left=of v] (i2);
\path (v)--(f1) coordinate[pos=0.5] (e);
\coordinate (h) at ($(e)+({1/sqrt(2)},-{1/sqrt(2)})$);
\diagram* {
(i1) --  (v) [blob] --  (i2),
(f2) --  (v),
(f1) --  (v),
};
\end{feynman}
\path (e)--++(-45:0.3) coordinate (e1);
\draw (e1) circle(0.3);
\end{tikzpicture}}
\quad
+
\quad
\resizebox{2.5cm}{!}{\begin{tikzpicture}[baseline={(current bounding box.center)}]
\begin{feynman}
\vertex (i1);
\vertex [left=2cm of i1] (f1);
\vertex [below =of i1] (i2);
\vertex [below=of f1] (f2);
\diagram* [horizontal=v1 to v2, layered layout] {
(i1) -- [half left] (f1),
(i2) -- [half right] (f2),
};
\end{feynman}
\end{tikzpicture}}
\quad
+
\quad
\cO(\lambda^3)
\ .
\end{dmath}

In both cases, some of the contributing Feynman diagrams are genuinely non-Gaussian, as the series expansions go beyond the  resummation of "cactus"-like diagrams carried out by a Gaussian ansatz. In the following subsection we will explain how the icMERA methods precisely accounts for such non-Gaussian contributions when a small coupling regime is considered. 

\subsection{Perturbative analysis of $G_c^{(n)}$ in icMERA}

Depending on the number of fields $n$ that we consider in the nonlinear canonical transformations presented in Section \ref{sec:nlct}, our icMERA variational ansatz will be more suitable to capture the ground state of the symmetric or asymmetric phases. For the $\lambda\phi^4$ potential, the transformation $\pi\phi^2$ (resp. $\pi\phi^3$) preserves (resp. breaks) the $\mathbb Z_2$ symmetry of the model and thus, approaches the ground state in the symmetric (resp. asymmetric) phase. 

Additionally, it is obvious that the type of transformation also affects the perturbative expansion of the correlators. In order to carry out an analysis of the genuine effects of the non-Gaussian transformations, we will denote the resummation of the cactus-type diagrams carried out by the Gaussian part of the ansatz, namely the purely Gaussian corrections to the propagator, by a thick line:
\begin{dmath}
\resizebox{3cm}{!}{\begin{tikzpicture}[baseline={(current bounding box.center)}]
\begin{feynman}
\diagram* {
a -- [very thick] b  -- [very thick] c  ,
};
\end{feynman}
\end{tikzpicture}}
\quad
=
\quad
\resizebox{3cm}{!}{\begin{tikzpicture}
\begin{feynman}
\diagram* {
a --  b  --  c ,
};
\end{feynman}
\end{tikzpicture}}
\quad
+
\quad
\resizebox{3cm}{!}{\begin{tikzpicture}
\begin{feynman}
\vertex (a);
\vertex [right=of a] (c);
\vertex [right=of c] (d);
\vertex [above=of c] (c1);
\vertex [above=of c1] (c2);
\diagram* {
(a) -- (c) -- (d) ,
};
\path (c)--++(90:0.3) coordinate (A);
\draw (A) circle(0.3);
\end{feynman}
\end{tikzpicture}}
\quad
+
\quad
\resizebox{3cm}{!}{\begin{tikzpicture}
\begin{feynman}
\vertex (a);
\vertex [right=of a] (c);
\vertex [right=of c] (d);
\vertex [above=.9 of c] (c1);
\diagram* {
(a) -- (c) -- (d) ,
};
\path (c)--++(90:0.3) coordinate (A);
\draw (A) circle(0.3);
\path (c1)--++(90:0) coordinate (A1);
\draw (A1) circle(0.3);
\end{feynman}
\end{tikzpicture}}
\quad
+
\quad
\cO(\lambda^3)
\ .
\end{dmath}

To implement the weak coupling limit in icMERA, we have to focus on the product $s \cdot c(\vp,\vq;u)$, which induces the ``variational vertices'' in the theory. Firstly, we have to impose that $s\propto \lambda$. This implies that, for the $G_c^{(k)}$ correlators, our formalism will capture $\cO(\lambda^k)$ Feynman diagrams.

Secondly, let us note that the optimization condition of icMERA for $f(\vp,\vq;u)$ in Eq. (\ref{eq:optimization}) of the main text, imposes a Schwinger-Dyson-like equation on the variational parameter $c(\vp,\vq;u)$, as it has been previously shown in Appendix B of \cite{JJ1} (see also \cite{ibanez}). Here, we observe that the leading order term of such vertex is proportional to the propagator, $c(\vp,\vq;u)\propto F(|\vp+\vq|;u)$.

As a consequence, the connected correlators \eqref{connected_corr} and, in particular, each of the $\chi_i$ integrals introduced in Appendix \ref{app:integrals} admit a precise diagrammatic expansion. Despite here we will restrict ourselves to the transformations $\pi\phi^2$ and $\pi\phi^3$, it would be interesting to study other nonlinear transformations and gain a better understanding of the generalized condensates that they give rise to.

\subsubsection*{The $\pi\phi^2$ entangler}

With this type of entangler, the scale dependent canonical transformation of the fields implemented by an icMERA circuit breaks the $\phi\to -\phi$ invariance of the $\lambda \phi^4$ model. Thus, in case the ground state is in the symmetric phase, the icMERA circuit uses the degrees of freedom in the symmetry-breaking transformation to minimize its energy \cite{polley89}. In the weak interacting limit, the connected 1-,2-,3-, and 4-pt functions are given by \cite{JJ1}
\begin{equation*}
\begin{aligned}
G_c^{(1)} 
&= 
\quad
\lambda\, \hspace{5pt}
\vcenter{\hbox{\begin{tikzpicture}
\begin{feynman}[small]
\vertex (v1) ;
\vertex [above right=of v1] (v2);
\vertex [right=0.5cm of v1] (c);
\vertex [below right=of v1] (v3);
\vertex[left=of v1] (i1);
\diagram* {
(i1) -- [very thick] (v1) 
,
};
\end{feynman}
\draw (c) [very thick] circle (5mm);
\end{tikzpicture}}}
\ ,
\\[15pt]
G_c^{(2)} 
&= 
\qquad 
\vcenter{\hbox{\begin{tikzpicture}
\begin{feynman}[small]
\diagram* {
a -- [very thick] b-- [very thick] c,
};
\end{feynman}
\end{tikzpicture}}}
&
\quad + \quad
&
\lambda^2\, \hspace{5pt}
\hbox{\begin{tikzpicture}
\begin{feynman}[small]
\diagram* {
a -- [very thick] b  -- [very thick]  c -- [very thick] d,
b  -- [half left,very thick] c ,
};
\end{feynman}
\end{tikzpicture}}
\ ,
\\[15pt]
G_c^{(3)} 
&= 
\qquad
\lambda\, \hspace{5pt}
\vcenter{\hbox{\begin{tikzpicture}
\begin{feynman}[small]
\vertex (v1) ;
\vertex [above right=of v1] (v2);
\vertex [below right=of v1] (v3);
\vertex[left=of v1] (i1);
\diagram* {
(i1) -- [very thick] (v1) -- [very thick] (v2),
(v1) -- [very thick] (v3),
};
\end{feynman}
\end{tikzpicture}}}
&\quad + \quad
&
\lambda^3\, \hspace{5pt}
\vcenter{\hbox{\begin{tikzpicture}
\begin{feynman}[small]
\vertex  (v1) ;
\vertex [above right=of v1] (v2) ;
\vertex [below right=of v1] (v3);
\vertex[left=of v1] (i1);
\vertex[right=of v2] (f1);
\vertex[right=of v3] (f2);
\diagram {
(v1) -- [very thick] (v2) -- [very thick] (v3) -- [very thick] (v1),
(v1) -- [very thick] (i1),
(v2) -- [very thick] (f1),
(v3) -- [very thick] (f2),
};
\end{feynman}
\end{tikzpicture}}}
\ ,
\\[15pt]
G_c^{(4)} 
&= 
\qquad
\lambda^2\, \hspace{5pt}
\vcenter{\hbox{\begin{tikzpicture}
\begin{feynman}[small]
\vertex (i1);
\vertex [ below right=of i1] (v1) ;
\vertex [ below left=of v1] (i2) ;
\vertex [right=of v1] (v2) ;
\vertex [above right=of v2] (f1);
\vertex [below right=of v2] (f2) ;
\diagram* {
(i1) --  [very thick](v1),
(v1) --  [very thick](i2),
(v1) --  [very thick](v2),
(f1) --  [very thick](v2),
(v2) --  [very thick](f2),
};
\end{feynman}
\end{tikzpicture}}}
&\quad + \quad
&
\lambda^4\, \hspace{5pt}
\vcenter{\hbox{\begin{tikzpicture}
   \begin{feynman}[small]
\vertex (vlu) ;
\vertex [right=of vlu] (vru) ;
\vertex [below=of vru] (vrd) ;
\vertex [left=of vrd] (vld) ;
\vertex[above left=of vlu] (i1) ;
\vertex[above right=of vru] (f1);
\vertex[below left=of vld] (i2);
\vertex[below right=of vrd] (f2);
\diagram* {
(vlu) -- [quarter left,very thick] (vru) -- [quarter left,very thick] (vrd) -- [quarter left,very thick] (vld) -- [quarter left,very thick] (vlu),
(i1) -- [very thick] (vlu),
(i2) -- [very thick] (vld),
(f1) -- [very thick] (vru),
(f2) -- [very thick] (vrd),
};
\end{feynman}
\end{tikzpicture}}}
\ .
  \end{aligned}
\end{equation*}
In order of appearance (except the "thick"-free propagator), each of these diagrams corresponds to the $\chi_i$ integrals introduced in Appendix \ref{app:integrals}, for $i=1,\ldots,6$.

\subsubsection*{The $\pi\phi^3$ entangler}

In this case, the $\mathbb{Z}_2$ symmetry of the model remains unbroken and only $G_c^{(2n)}$ are nonzero. As a consequence, the non-Gaussian diagrams in \eqref{eq:2p-feynman} and \eqref{eq:4p-feynman} are captured.  In this case, the $G_c^{(2)}$ and $G_c^{(4)}$ correlators admit the following expansion:
\begin{equation*}
\begin{aligned}
G_c^{(2)} 
&
\quad = \quad 
\vcenter{\hbox{\begin{tikzpicture}[baseline={(current bounding box.center)}]
\begin{feynman}
\diagram* {
a -- [very thick] b-- [very thick] c,
};
\end{feynman}
\end{tikzpicture}}}
\quad + \quad
\lambda^2\, \hspace{5pt}
\hbox{\begin{tikzpicture}[baseline={(current bounding box.center)},scale=7]
\begin{feynman}[small]
\diagram* {
a --[very thick]  b   --[very thick]  c --[very thick]  d  --[very thick]  e,
};
\draw (c) [very thick] circle (1.4mm);
\end{feynman}
\end{tikzpicture}}
\ ,
\\[15pt]
G_c^{(4)} 
&
\quad = \quad
\lambda\, \hspace{5pt}
\vcenter{\hbox{\begin{tikzpicture}
\begin{feynman}
\vertex (v);
\vertex [above left=of v] (i1);
\vertex [below left=of v] (i2);
\vertex [above right=of v] (f1);
\vertex [below right=of v] (f2);
\diagram* [horizontal=v1 to i1, layered layout] {
(i1) --[very thick] (v) --[very thick] (i2) --[very thick] (v) --[very thick] (f1) --[very thick] (v) --[very thick] (f2),
};
\end{feynman}
\end{tikzpicture}}}
\quad + \quad
\lambda^2\, \hspace{5pt}
\vcenter{\hbox{\begin{tikzpicture}
\begin{feynman}
\vertex (i1);
\vertex [left= 3cm of i1] (f1);
\vertex [below=2cm of i1] (i2);
\vertex [below=2cm of f1] (f2);
\diagram* [horizontal=v1 to v2, layered layout] {
(i1) --[very thick,half left] (f1),
(i2) --[very thick,half right] (f2),
};
\end{feynman}
\end{tikzpicture}}}
\quad + \quad
\lambda^3\, \hspace{5pt}
\vcenter{\hbox{\begin{tikzpicture}
\begin{feynman}
\vertex (c);
\vertex [left=of c](v1);
\vertex [above=1cm of c] (v2);
\vertex [below=1cm of c] (v3);
\vertex[above left=of v1] (i1);
\vertex[below left=of v1] (i2);
\vertex[right=1cm of v2] (f1);
\vertex[right=1cm of v3] (f2);
\diagram* {
(v1) --[very thick]  (v2) --[very thick]  (v3) --[very thick]  (v1),
(i2) --[very thick] (v1) --[very thick]  (i1),
(v2) --[very thick]  (f1),
(v3) --[very thick]  (f2),
(v2) --[very thick, quarter left] (v3),
};
\end{feynman}
\end{tikzpicture}}}
\\[15pt]
&\qquad\qquad + \quad
\lambda^4\, \hspace{5pt}
\vcenter{\hbox{\begin{tikzpicture}
  \begin{feynman}[small]
    \vertex (a);
    \vertex [below      =of a] (b);
    \vertex [above left=of a] (i1) ;
    \vertex [below left =of b] (i2) ;
    \vertex [right      =of a] (ar) ;
    \vertex [right      =of b] (br) ;
    \vertex [above right=of ar] (f1);
    \vertex [below right=of br] (f2);

    \diagram* {
      (i1) --[very thick] (a),
      (i2) --[very thick] (b),
      (f1) --[very thick] (ar),
      (f2) --[very thick] (br),
      (a) --[very thick] (br),
 };
  \end{feynman}
\draw [link] (b) -- (ar);
\path (a)--(br) coordinate[pos=0.5] (c);
\draw (c) [very thick] circle (.75cm);
\end{tikzpicture}}}
\quad + \quad
\lambda^4\, \hspace{5pt}
\vcenter{\hbox{\begin{tikzpicture}
  \begin{feynman}[small]
    \vertex (a);
    \vertex [below      =of a] (b);
    \vertex [above left =of a] (i1) ;
    \vertex [below left =of b] (i2) ;
    \vertex [right      =2cm of a] (ar) ;
    \vertex [right      =2cm of b] (br) ;
    \vertex [above right=of ar] (f1);
    \vertex [below right=of br] (f2);

    \diagram* {
      (ar) -- [very thick,quarter right] (b),
      (ar) -- [very thick,quarter left] (b),
      (a) -- [link, very thick,quarter right] (br),
      (a) -- [link,very thick,quarter left] (br),
      (a) -- [very thick,quarter right] (br),
      (a) -- [very thick,quarter left] (br),
      (a) -- [very thick,quarter left] (ar),
      (b) -- [very thick,quarter right] (br),
      (i1) --[very thick] (a),
      (i2) --[very thick] (b),
      (f1) --[very thick] (ar),
      (f2) --[very thick] (br),
      };
  \end{feynman}
\end{tikzpicture}}}
\ .
  \end{aligned}
\end{equation*}
Further details on the analog of the $\chi_i$ generalized condensates for the $\pi\phi^3$ transformation can be found in \cite{ritschel90}.

\section{$\chi$ integrals}
\label{app:integrals}
The loop integrals $\chi_i$ which are related to the circuit $\pi\, \phi^2$ depend on both positions and the renormalization scale $u$. Once the optimal variational parameters $f(p;u)$ and $f(p,q_1,q_2;u)$ are obtained for a particular theory, then higher order correlation functions can be computed through them. Denoting
\begin{align}
c(\vp,\vq;u)
\equiv \tilde h_u (|\vp+\vq|e^{-u},pe^{-u},qe^{-u})\, 
,
\end{align}
their explicit expressions can be written as 
\begin{align}
\label{eq:chis}
 \chi_1(u)
=&\
\frac12 \int_\vp c (\vp,-\vp;u) \ F(pe^{-u})
\ ,
\\ \nonumber
\\
 \chi_2(\va\vb;u) 
=&\
\frac12 \int_{\vp\vq}  \, c (\vp,\vq;u)^2 F(p e^{-u})F(q e^{-u}) \, e^{i(\vp+\vq)\cdot \vx_{ab}} 
\ ,
\\ \nonumber
\\ 
 \chi_3(\va\vb,\vc\vd;u) 
=&\
\frac12 \int_{\vp\vq} \, c (\vp,\vq;u)\, F(p e^{-u})F(q e^{-u})\, e^{i(\vp\cdot \vx_{\va\vb}+\vq\cdot \vx_{\vc\vd})}
\ ,
\\ \nonumber
\\ 
\chi_4(\va\vb,\vc\vd,\ve\vf;u) 
=&\
\int_{\vp\vq\vr} \, c (\vp ,\vq;u)\,  c (\vq ,-\vr;u)\,  c (\vp ,\vr;u) F(p e^{-u})F(q e^{-u})F(r e^{-u})\\ \nonumber
&\times e^{i(\vp\cdot \vx_{\va\vb}+\vq\cdot \vx_{\vc\vd}+\vr\cdot \vx_{\ve\vf})}
\ ,
\\ \nonumber
\\
\vspace{5pt}
\chi_5(\va\vb,\vc\vd,\ve\vf;u)  
=&\
\int_{\vp\vq\vr} \, c (\vp ,\vr;u)\,   c (\vq ,\vr;u)
F(p e^{-u})F(q e^{-u})F(r e^{-u}) e^{i(\vp\cdot \vx_{\va\vb}+\vq\cdot \vx_{\vc\vd}+\vr\cdot \vx_{\ve\vf})}
\ ,
\\ \nonumber
\\
\vspace{10pt}
\chi_6(\va\vb,\vc\vd,\ve\vf,\vg\vh;u)  
=&\
\int_{\vp\vq\vr\vs}  c (\vp ,\vq;u)\, c (\vp ,\vr;u)\,  c (\vq ,\vs;u)\, c (\vr ,\vs;u) \\ \nonumber
&\times F(p e^{-u})F(q e^{-u})F(r e^{-u})F(s e^{-u}) e^{i(\vp\cdot \vx_{\va\vb}+\vq\cdot \vx_{\vc\vd}+\vr\cdot \vx_{\ve\vf}+\vs\cdot \vx_{\vg\vh})}
\ ,
\\ \nonumber
\\
\chi_7(u)
=&\
\frac14 \int_{\vp\vq} \Big[c(\vp,\vq;u)^2 (|\vp+\vq|e^{-u})^2 F(pe^{-u})\, F(qe^{-u})\\ \nonumber
&+  c(|\vp+\vq|,-\vq;u)^2 F(qe^{-u})\, F(pe^{-u})^{-1}
\Big]\, .
\end{align}

With this, the quantities in brackets appearing in \eqref{connected_corr} are given by
\begin{align}
[\chi_5(1234)]
=&\
\chi_5(12,32,14)
+\chi_5(12,42,13)
+\chi_5(13,23,14)
+\chi_5(13,43,12)
\nonumber\\
&\
+\chi_5(14,24,13)
+\chi_5(14,34,12)
+\chi_5(23,13,24)
+\chi_5(23,43,21)
\nonumber\\
&\
+\chi_5(24,14,23)
+\chi_5(24,34,21)
+\chi_5(34,14,32)
+\chi_5(34,24,31)\ ,\nonumber\\[5pt]
[\chi_6(1234)] 
=&\
\chi_6(12,23,34,14)
+\chi_6(13,34,24,12)
+\chi_6(13,14,23,24)\ ,
\end{align}
where the explicit dependence on $u$ has been dropped for clarity.


\end{document}